\documentclass[%
reprint, 
superscriptaddress,
amsmath,amssymb,aps,
pra,
floatfix
]{revtex4-2}
\usepackage{float}
\usepackage{physics}
\usepackage{comment}
\usepackage{xcolor}
\usepackage{graphicx}
\usepackage{dcolumn}
\usepackage{bm}
\usepackage{cleveref}
\usepackage{graphicx}
\usepackage{tabularx}
\usepackage[labelformat=simple]{subcaption}

\begin{document}

\preprint{APS/123-QED}

\title{Non-Markovian Dynamics of a Single Excitation within Many-Body Dissipative Systems}

\author{Adam Burgess}
\email{a.d.burgess@surrey.ac.uk}
 \affiliation{Leverhulme Quantum Biology Doctoral Training Centre, University of Surrey, Guildford, GU2 7XH, United Kingdom}
\affiliation{Advanced Technology Institute,  University of Surrey, Guildford, GU2 7XH, United Kingdom}
\affiliation{Department of Physics,  University of Surrey, Guildford, GU2 7XH, United Kingdom}
\author{Marian Florescu}%
\email{m.florescu@surrey.ac.uk}
\affiliation{Advanced Technology Institute,  University of Surrey, Guildford, GU2 7XH, United Kingdom}
\affiliation{Department of Physics, University of Surrey, Guildford, GU2 7XH, United Kingdom}


\date{\today}

\begin{abstract}
We explore the dynamics of $N$ coupled atoms to a generic bosonic reservoir under specific system symmetries. In the regime of multiple atoms coupled to a single reservoir with identical couplings,  we identify remarkable effects, notably that the initial configuration of the atomic excited state amplitudes strongly impacts the dynamics of the system and can even fully sever the system from its environment. Additionally, we find that steady state amplitudes of the excited states become independent of the choice of the reservoir.  The framework introduced is applied to a structured photonic reservoir associated with a photonic crystal, where we show it reproduces previous theoretical and experimental results and it predicts superradiant behaviour within the single-excitation regime.
\end{abstract}

\maketitle

\section{Introduction}
Understanding the dynamics of atomic systems within dissipative environments has played a pivotal role in developing new and better artificial structures that have the capacity for processing information on shorter spatial and temporal scales ~\cite{CoherentSCQuantumDots,GateOperationQubits,SimulStateMeasureJosephsonQB}. The conventional approach to studying such systems is to deploy the theory of open quantum systems~\cite{TheoryOQSBook}, wherein the archetypal example is spin systems coupled to a bosonic reservoir are utilised to model a two-level atomic system coupled to a quantised electromagnetic field. In principle, unravelling the combined dynamics of the two systems is exceedingly difficult and often it is merely the dynamics of the atomic system that is of interest. Therefore, a scheme is required to reduce out the environmental degrees of freedom. A common approach to achieving this is by invoking the Markov approximation, which is valid for environments that recover instantaneously from interacting with the system ~\cite{BornMarkovApproxAtomLaser}. Although this approach has proven fruitful in understanding many systems it fails to adequately capture quantum induced memory effects in the system. For example, within micro-structured photonic systems, such as photonic crystals, the local density of states for the electromagnetic field varies rapidly near to the band edges of the photonic band gap. Such rapid fluctuations make the Markov approximation invalid ~\cite{burgess_modelling_2021,SingleAtomSwitch,john_florescu_2001}, and embedding a two-level atomic system with its transition energy near to the band edge of a photonic band gap results in highly non-Markovian effects. The strong interaction between the atomic system and the photonic reservoir leads to a dressing of the atomic states highly intertwining the attributes of the photonic reservoir and the atomic degrees of freedom. The corresponding effects lead to temporal oscillations and fractional decay of atomic population, spectral splitting and sub-natural line-widths of the atomic transitions~\cite{John1994}.
Furthermore, as we begin to scale up the technology of these artificial systems, it becomes  necessary to understand the collective effects as the number of two-level systems coupled to the reservoir is increased and understanding the control mechanisms behind this may lead to better insights in designing such systems which may be used in the implementation of quantum networks~\cite{KimbleQuantumNetwork} and clocks~\cite{QuantumNetworkClocks}. 
In this work we study the single-excitation regime of many-body dissipative systems for generic reservoirs and reveal some remarkable characteristics. We also apply the formalism developed for a model system, of the isotropic photonic crystal, a particularly interesting example which displays what can be considered the highest degree of non-Markovianity for a reservoir due to the divergence in its density of states at the photonic band edge~\cite{john_florescu_2001}. Such spin-coupled systems have been realised in a plethora of experiments, including quantum dots~\cite{quantumdotspincouple}, superconducting networks~\cite{Superconducting}, trapped ions~\cite{trappedion},  cold-atoms inside of optical lattices~\cite{OpticalLattice} and more recently within photonic crystal wave guides~\cite{KimbleOPhCr,Yu12743,SinglePhotonQD}.

Previous studies have considered the two-qubit interaction with independent reservoirs utilising the second order time-convolutionless approach~\cite{NM2Qubit}, the study of entanglement of two-qubits coupled to a single reservoir~\cite{NM2qubitSE}, the single-excitation dynamics for a single-spin in a non-Markovian reservoir~\cite{Breuer_1999} and the single-excitation dynamics of two coupled atoms inside three dimensional anisotropic photonic crystals~\cite{shen_quantum_2016}. In this study we  go further by studying the dynamics of $N$-atoms inside of generic bosonic reservoirs that can be adequately described by a local density of states around the atomic system location. Such models are useful in understanding the induced dynamics of atomic systems in a range of environments such as the quantised electromagnetic fields as well as non-zero temperature vibronic environments.

This article is organised as follows: In Sec.~2 we employ the Heisenberg equations of motion to study the single excitation dynamics of $N$ coupled atomic systems embedded within a single bosonic reservoir. Two system topologies are considered, the first being symmetric coupling wherein all the atoms are coupled to one another with the same coupling strength and a nearest neighbour system wherein atoms only interact with adjacent partners describing a chain of atoms. We study the late time behaviour of the exited state populations for each system. In Sec.~3 we contrast these results with those for independent reservoir dissipative systems. Finally, in Sec.~4 we consider the specific case of the photonic crystal and derive the full time evolution for the system wherein the transition energies of the atoms are close to an isotropic photonic band gap - a model system that has very strong non-Markovian characteristics due to a divergence in local density of states of the electromagnetic field around a photonic band gap.

\section{Many Atoms in a Single Bosonic Reservoir}
We begin with a system of $N$ two-level atoms embedded within a single structured reservoir, with all atoms coupling to the reservoir with the same coupling strengths. To facilitate analytic solutions we assume that the reservoir is initially in its vacuum state, and we restrict our study to the single-photon excitation regime. Such assumptions are justified for very low temperature systems. Furthermore, in the case of photonic crystals with a photonic band edge energetic modes of the electromagnetic field with energy less than the band gap frequency $\omega_C$ are not accessible and as such cannot be excited by thermal fluctuations. For band gap frequencies in the optical range the thermal fluctuations exciting electromagnetic field modes with energy greater than the band gap are negligible~\cite{Existence,FCC}.  Furthermore, we deploy the rotating wave approximation (RWA) in order to derive analytical solutions.

\subsection{Fully Symmetric Coupling}
We begin by considering a many-body dissipative system wherein the atomic systems are coupled to each other and to the bosonic reservoir with the same coupling strengths. For such a model, there are two relevant length scales, the first the spatial range of the inter-atomic coupling generated by the dipole-dipole interaction. The second is the total size of the atomic ensemble. We require the local density of states for the reservoir across both length scales to remain the same. However, it is clear that the length scale of the dipole-dipole interaction is necessarily smaller than the ensemble size and such models have been used to effectively describe super-radiant effects in electromagnetic field reservoirs~\cite{SuperRadiancePBG,Superradiance,Dicke}. The system Hamiltonian in the rotating wave approximation Hamiltonian is given by
\begin{align}
    H = &\sum_{i=1}^N\omega_0\sigma_i^+\sigma_i^- + \sum_\lambda \omega_\lambda a^\dag_\lambda a_\lambda +  \nonumber \\&i\sum_{i,\lambda}g_\lambda(a_\lambda \sigma^+_i -a^\dag_\lambda \sigma^-_i ) + \sum_{i\neq j}J\sigma^+_i\sigma_j^-
\end{align}
where $\sigma_i^+$and $\sigma_i^-$, are the excitation and de-excitation operators for the $i$th atomic system respectively, $a_{\lambda}$ and $a^{\dagger}_{\lambda}$ are the bosonic field annihilation and creation operators, $\omega_0$ and $\omega_{\lambda}$ are the atomic transition and the $\lambda$-boson mode frequencies,  $g_\lambda$ is the coupling strength of the atomic system and the $\lambda$-boson mode~\cite{Pfeifer,AllenEberly,SuperRadiancePBG} and $J$  is the dipole-dipole coupling strength between atoms. This is equivalent to a Dicke model~\cite{Dicke} for $N$-atoms with a coupling between each of the atoms. 

The Hamiltonian has the convenient property that it conserves the excitation number~\cite{Garraway}, that is to say it commutes with the number operator $N = \sum_{i=1}^N\sigma_i^+\sigma_i^- + \sum_\lambda a_\lambda^\dag a_\lambda$. As such if we consider the single excitation wavefunction given by 
\begin{equation}
    \phi(0) = c_0\psi_0+\sum_i^N c_i(0)\psi_i +\sum_\lambda c_\lambda(0) \psi_\lambda,
\end{equation}
where $\psi_i = \ket{i}_A\ket{0}_B$ is the state wherein the $i$th atom is in its excited state and all other atoms and the reservoir are in the ground state. $\psi_\lambda = \ket{0}_A\ket{\lambda}_B$ represents all atoms in their ground state and the bosonic system has its $\lambda$ mode excited. $\psi_0 = \ket{0}_A\ket{0}_B$ denotes the ground state of the entire system.
The time evolution of this state is given by
\begin{equation}
    \phi(t) = c_0\psi_0 + \sum_i^N c_i(t)\psi_i +\sum_\lambda c_\lambda(t) \psi_\lambda.
\end{equation}
It is convenient to introduce the following parameter that sums over the excited state amplitudes,
\begin{equation}
    c_+(t) = \sum^N_{i=1} c_i(t).
\end{equation}

The Heisenberg equations of motion  for the state amplitudes read:
\begin{align}
    \Dot{c}_i =& -iJ(c_+-c_i) + \sum_\lambda c_\lambda g_\lambda e^{i(\omega_0-\omega_\lambda)t},\nonumber \\
    \Dot{c}_+ =& -iJ(N-1)c_+ + N \sum_\lambda c_\lambda g_\lambda e^{i(\omega_0-\omega_\lambda)t}, \nonumber \\
    \Dot{c}_\lambda =& -g_\lambda e^{-i(\omega_0-\omega_\lambda)t} c_+.
    \label{HEOM1}
\end{align}

Assuming now that the bosonic field is initially in its vacuum configuration ($c_\lambda(0) = 0 , \forall\,\lambda$), formally integrating the Heisenberg equations of motion yields
\begin{equation}
    c_\lambda(t) = -\int^t_0 g_\lambda e^{-(\omega_0-\omega_\lambda)t_1}c_+(t_1)dt_1.
    \label{eq_c_lambda}
\end{equation}

Introducing the so called  `memory kernel' $G(t) = \sum_\lambda g_\lambda^2 e^{i(\omega_0-\omega_\lambda)t}$ and substituting  Eq.~\ref{eq_c_lambda} in Eqs.~ \ref{HEOM1} we obtain
\begin{align}
    \Dot{c}_i =& -iJ(c_+-c_i) - \int^t_0G(t-t_1)c_+(t_1)dt_1,\nonumber \\
    \Dot{c}_+ =& -iJ(N-1)c_+ - N\int^t_0G(t-t_1)c_+(t_1)dt_1.
    \label{eq_dot_c}
\end{align}

Note that the dynamics induced by the bosonic reservoir is solely controlled by the value of $c_+$ convoluted with the memory kernel $G(t)$; this can be interpreted as the reservoir only coupling to the total polarisation of the collection of atoms. To emphasise this point, we can consider the total polarisation operator $\sigma^-_T = \sum_{i=1}^N \sigma_i^-$,  that is the sum of the polarisation for each of the atomic systems. The expectation value of this operator is $\langle\sigma^-_T\rangle = c_0c_+(t) \propto c_+(t)$, suggesting that $c_+$ is a measure of the total systems polarisation. Similarly, the atom-atom coupling term is  $\propto c_+ - c_i$, so  we can interpret this as the individual atom coupling to the aggregation of all the other atoms total polarisation. Additionally, it is this convolution term in Eq.~\ref{eq_dot_c} that controls the non-Markovianity of our dynamics as it integrating over all previous states of the system. 

Eq.~\ref{HEOM1} can be solved in terms of the Laplace transform of the single atom amplitudes $c_i$ and $c_+$ (the full derivation are presented in Appendix A1).  
For the value $c_+$ corresponding to the total polarisation of the atoms we obtain 
\begin{equation}
    \Tilde{c}_+(s) = \frac{c_+(0)}{s+iJ(N-1)+\Tilde{G}(s)N},
    \label{eq_c_plus_s}
\end{equation}
where $\Tilde{G}(s)$ is the Laplace transform of the memory kernel. We can relate $\Tilde{G}(s)$  to the spectral density $J(\omega)$ - describing the coupling strengths of the atomic systems to the environment at different environmental frequencies defined by
\begin{equation}
    J(\omega)= \sum_\lambda g_\lambda^2 \delta(\omega-\omega_\lambda).
\end{equation}
Thus the relation between the Laplace transform of the memory kernel and the spectral density is
\begin{equation}
\Tilde{G}(s) =\int d\omega \frac{J(\omega)}{s-i(\omega_0-\omega)}.
\end{equation}
Interestingly the dynamics associated with the total polarisation parameter $c_+$ maps onto the single atomic system 
\begin{equation}
    \Tilde{c}_1(s) = \frac{c_1(0)}{s+\Tilde{G}(s)},
       \label{eqn:FCc1s}
\end{equation}
with rescaled coupling strengths $g_\lambda \rightarrow \sqrt{N}g_\lambda$ and a Lamb shift of the transition frequency $\omega_0 \rightarrow \omega_0 +J(N-1)$, picking up a phase parameter $e^{iJ(N-1)t}$ in the time domain. The Laplace transforms of the individual state  amplitudes are given by
\begin{equation}
    \Tilde{c}_i(s) = \frac{c_i(0)}{s-iJ} - \frac{c_+(0)(\Tilde{G}(s)+iJ)}{(s-iJ)(s+iJ(N-1)+N\Tilde{G}(s))}.
    \label{eqn:FCcis}
\end{equation}

The Laplace transform solutions in Eqs.~\ref{eqn:FCc1s}-\ref{eqn:FCcis} provide a few interesting results as we note that the coupling to the reservoir dynamics is strongly dependent on the initial value of the total polarisation $c_+(0)$. We also note that we can decouple the atoms from their environment by choosing an appropriate initial condition such that $c_+(0)=0$. Doing so leaves only the first term on the right hand side of Eqn.\ref{eqn:FCcis} non-zero. The inverse Laplace transform of this leads to only time evolution in the phase of each excited state amplitude $c_i(t)=c_i(0)e^{iJt}$ and does not affect population dynamics. This is because the total polarisation for the system is zero preventing the atoms from coupling to the environment. Another initial condition of note is full initial symmetry, such that each atom has equivalent time evolution. Each atom acts identically towards the evolution of the total polarisation and as such each acts like a single atom within a dissipative environment as the gain in excitation due to energy transfer between atoms is balanced with the loss. For example, assuming $c_i(0)=\frac{1}{\sqrt{N}}$ and $c_+(0)=\sqrt{N}$ the evolution of the excited state amplitudes is governed by  
\begin{equation}
    \Tilde{c}_i(s) = \frac{1}{\sqrt{N}(s+iJ(N-1)+N\Tilde{G}(s))},
\end{equation}
which is simply the re-scaled dynamics of $c_+$ in Eq.~\ref{eq_c_plus_s}. 

We find also for the steady state of these systems is non-zero as the total polarisation for the system tends to relax, this however, does not necessitate that individual atom relax back to the ground state. Using the Final Value Theorem (FVT) and removing the phase dependence ($s\to s+iJ$), yields 
\begin{equation}
    c_{i\infty} = \lim_{s\to 0} s\Tilde{c}_i(s+iJ) = c_i(0) -\frac{c_+(0)}{N},
    \label{eqn:ciss}
\end{equation}
so we expect the late time population $\abs{c_i}^2$ to tend towards $|c_i(0) -\frac{c_+(0)}{N}|^2$ minus oscillations as the FVT does not account for these. Effectively, by configuring the initial condition of the atomic systems, we can localise excitations in particular atoms. Such a technique may have relevance in quantum memory storing devices. 

An interesting case occurs when the total polarisation for the atoms is equal to a single atom's polarisation ($c_1(0) = c_+(0)$). As the environment only couples to the total polarisation and the relaxation of the total polarisation is carried equally across all of the atomic systems this allows for non-zero steady state polarisations for individual atoms even in highly dissipative or Markovian environments - where the total polarisation goes to zero. This is apparent when we apply the final value theorem with this initial condition yielding $c_{1\infty} = c_1(0)(1-\frac{1}{N})$. As we increase the number of atoms within the system we have $\lim_{N\to \infty} c_{1\infty} = c_1(0)$ so in the late time we return back to our initial state value (minus phase contributions). In fact scaling the numbers of atoms effectively increases the speed of the relaxation of the total polarisation as its coupling to the reservoir $\sqrt{N}g_\lambda$ is dependent on $N$, so we can imagine that as $N$ increases we actually get limited reservoir dynamics induced on the individual atoms as the changes in the total polarisation are spread amongst all $N$ atoms.

We now present an effective Hamiltonian model for a single atomic system to show how the dynamics we are seeing are similar to that of single atomic system with an altered effective Hamiltonian. To model this behaviour we consider a single atomic system embedded within a bosonic reservoir with an additional self-interaction term to account for the atom-atom interactions. The system Hamiltonian is given by 
\begin{align}
    \Tilde{H} =& \omega_0\sigma_+\sigma_- +\sum_\lambda \omega_\lambda a^\dag_\lambda a_\lambda  +  J(N-1) \sigma^+\sigma^- + \\&+i\sqrt{N}\sum_{\lambda}g_\lambda(a_\lambda \sigma^+ e^{i(\omega_0-\omega_\lambda)t} -a^\dag_\lambda \sigma^- e^{-i(\omega_0-\omega_\lambda)t}).
\end{align}
We note that the coupling to the reservoir of this new aggregated atom is stronger (for $N>1$) than in the original system. This leads to an interaction Hamiltonian of the form
\begin{align}
    \Tilde{H}_I =& J(N-1) \sigma^+\sigma^- \\&+i\sqrt{N}\sum_{\lambda}g_\lambda(a_\lambda \sigma^+ e^{i(\omega_0-\omega_\lambda)t} -a^\dag_\lambda \sigma^- e^{-i(\omega_0-\omega_\lambda)t}),
\end{align}
and the dynamical equation for the excited state amplitude $c'_+$
is given by 
\begin{equation}
     \Dot{c}'_+ = -iJ(N-1) c'_+ + \sqrt{N}\sum_\lambda c_\lambda g_\lambda e^{i(\omega_0-\omega_\lambda)t},
\end{equation}
yielding 
\begin{equation}
    \Dot{c}'_+ = -iJ(N-1)c'_+ - N\int^t_0G(t-t_1)c'_+(t_1)dt_1.
\end{equation}
which is equivalent to Eqn.\ref{eq_dot_c}. As such we can consider that the reservoir-atom interaction is being mediated by the aggregated system that has internal couplings that the reservoir cannot see. 

Now we consider how altering the coupling $J$ between the atoms impacts the steady state dynamics. In the regime where $J\rightarrow 0$ the steady state behaviour - non-zero excited state populations - still occurs, providing that 
\begin{equation}
\lim_{s\rightarrow 0} \frac{\Tilde{G}(s) }{s+N\Tilde{G}(s)}
\end{equation}
exists. However, the Laplace transform of the memory kernel in terms of the spectral density is given by 
\begin{equation}
    \lim_{s\to 0^+}  \Tilde{G}(s) = \lim_{s\to 0^+}\int d\omega \frac{J(\omega)}{s-i(\omega_0-\omega)}.
\end{equation}
Utilising the Sokhotski–Plemelj theorem~\cite{plemelj1908erganzungssatz} we obtain 
\begin{equation}
    \lim_{s\to 0^+}  \Tilde{G}(s) = J(\omega_0)\pi - i \mathcal{P}\int_0^\infty d\omega \frac{J(\omega)}{\omega-\omega_0},
\end{equation}
where $\mathcal{P}$ refers to the Cauchy principal value. As such, we expect the above to be generally non-zero and thus that we have non-zero steady state values.  For a generic Ohmic type spectral density with Ohmic parameter $p$ given by
\begin{equation}
    J(\omega) =\lambda \Omega \left[\frac{\omega}{\Omega}\right]^p e^{-\omega/\Omega},
\end{equation}
where $\lambda$ is the coupling strength and $\Omega$ the cutoff frequency. 
which leads to 
\begin{align}
    \lim_{s\to 0^+}  \Tilde{G}(s) =& J(\omega_0)[\pi(1 + i\text{Cot}(\pi p) -i(-1)^p\text{Csc}(\pi p))    \nonumber\\&- 
   i (-1)^p \Gamma(1 + p) \Gamma(-p, -\omega_0/\Omega)],
\end{align}
where $\Gamma(\cdot)$ and $\Gamma(\cdot,\cdot)$ are the gamma~\cite{Gamma} and incomplete gamma functions~\cite{IncompleteGamma},  respectively.
As this value of the Laplace transform for the memory kernel is non-zero we do not have issues in reducing out this term in finding the final value for the excited state amplitudes. Thus, reservoirs that can be modelled by the Ohmic types of spectral density would yield the steady states predicted by Eqn.\ref{eqn:ciss}.

\subsection{Nearest Neighbour Coupling}
In this section we consider a different model for atomic system in dissipative environments. Here we consider an atomic chain and allow the atoms to only interact directly with next neighbouring  atoms. Such a model is interesting as we are no longer limited by the number of dimensions we are embedding the system and hence making it a more viable experimental setup~\cite{KimbleSuperradiance}. This model is also much less restrictive in terms of the symmetry of the dipole-dipole coupling strength $J$ as we need only to assume that atoms are equidistant from their nearest neighbours. However, more care is needed in justifying coupling to the same reservoir, i.e that the $g_\lambda$ are shared across all atomic systems. However, note that dipole-dipole interactions occur over short length scales of less than $10$nm~\cite{dipole-dipole}, naturally we would not expect the electromagnetic field to vary greatly over this length scale justifying the identical coupling. Alternatively, if considering coupling to a photonic crystal reservoir, the atoms coupled be place at spatially equivalent positions in the crystalline structure.
The system Hamiltonian in the interaction picture is given by
\begin{align}
    \Tilde{H}_I = &i\sum_{i,\lambda}g_\lambda(a_\lambda\sigma^+_i e^{i(\omega_0 -\omega_\lambda)t} - a^\dag_\lambda\sigma^-_i e^{-i(\omega_0 -\omega_\lambda)t}) +\nonumber \\ &\sum_{j=1}^{N-1}J(\sigma_j^+\sigma_{j+1}^- + \sigma_j^-\sigma_{j+1}^+),
\end{align}
where we note that, as stated above,  that only next neighbouring atoms are coupled to each other.
By performing a similar analysis to the the highly symmetric case and keeping only terms up to second order in the dipole-dipole coupling parameter $J$,   we obtain  the following equation for excited state amplitude of the first atom in the chain
\begin{align}
    \Tilde{c}_1(s) =& \frac{s^2c_1(0)-iJsc_2(0)-J^2c_3(0)}{(s^2+J^2)s}\nonumber \\
    &+\frac{\Tilde{c}_+(s)\Tilde{G}(s)(J^2+iJs-s^2)}{(s^2+J^2)s}.
\end{align}
We can derive a similar expression for the Laplace transform solution for the excited state amplitude of the final atom in the chain
\begin{align}
    \Tilde{c}_N(s) = & \frac{s^2c_N(0)-iJsc_{N-1}(0)-J^2c_{N-2}(0)}{(s^2+J^2)s}\nonumber\\&+\frac{\Tilde{c}_+(s)\Tilde{G}(s)(J^2+iJs-s^2)}{(s^2+J^2)s}.
    \label{eqn:NNcN}
\end{align}
Keeping terms up to second order in the dipole-dipole coupling $J$ effectively allows for the dynamics of the nearest two atoms to intervene directly in the systems dynamics. We note that the reservoir induced dynamics (terms proportional to $\Tilde{G}(s)$) are again determined by the total polarisation parameter $c_+$.  However, as the atoms are only coupled to their nearest neighbours,  we pick up additional contributions from the dipole-dipole coupling $J$.

In contrast to the fully symmetric case we now pick up oscillatory terms given by the initial conditions of each of the two nearest atoms as shown by the terms independent of $\Tilde{G}(s)$ (these are simple Laplace transform identities for integrals of the trigonometric sin and cos functions). This demonstrates  the clear bi-directionality of the energy transfer that was not present in the previous model.

For convenience we now consider the expansion up to 1st order in $J$ and assume only that the first atom is initially excited,  $c_1(0) =1$ and all other atoms are initially in the round state,  $c_i(0)=0$. This yields 
\begin{align}
    \Tilde{c}_1(s) &= \frac{1}{s} - \frac{\Tilde{G}(s)}{(s^2 + N\Tilde{G}(s)s -2iJ\Tilde{G}(s)+ 2iJs)}\nonumber\\
    \Tilde{c}_N(s) &= - \frac{\Tilde{G}(s)}{(s^2 + N\Tilde{G}(s)s -2iJ\Tilde{G}(s)+ 2iJs)}.
\end{align}
In order to explore the late time dynamics,  we consider the transformation $s\rightarrow s+\frac{2iJ}{N}$. Up to first order in the dipole-dipole coupling $J$, we have 
\begin{equation}
    c_1(\infty) = 1 - \frac{\Tilde{G}}{2iJ(1+\frac{2}{N})+N\Tilde{G}}.
\end{equation}
Unlike the fully symmetric case explored previously, the late time dynamics appears to be governed by the value of $\Tilde{G}(0)$. However, for large values of $\Tilde{G}(0)$ or large values of $N$ and with $\Tilde{G}(0)\neq 0$ we recover similar dynamics to the fully symmetric system presented above.  Conversely, for smaller values of $N$ we find that the steady state character is strongly determined by the relation between the dipole-dipole coupling strength $J$ and the memory kernel $\Tilde{G}(s)$ . In Appendix A3, we show that even for small perturbations of the transition energy in the first and final atom in the chain we recover the same dynamics with a phase shift and frequency shift in these modified atoms' amplitudes. Such a correction shows that for small perturbations in these transition energies we can retain the dynamics of the combined system ensuring the system is robust. 

\section{Many Atoms in Separate Bosonic Reservoirs}
Here we consider the case wherein each of the atoms are coupled to independent bosonic reservoirs. This model assumes that the local environment between atomic systems has varied sufficiently that there are no correlations between the local environments of different atoms. Similar models are utilised in quantum chemistry to model the energy transfer in light harvesting complexes~\cite{LHC}, and we explore it here in our formalism merely to contrast it with the single reservoir case.
\subsection{Fully Symmetric Coupling}
For independent reservoirs interacting with each of the atoms we need to introduce separate creation and annihilation operators for the different reservoirs. The interaction Hamiltonian in the Dirac picture in this regime then becomes
\begin{align}
    \Tilde{H}_I &= i\sum_{i\lambda}g_{\lambda}(a_{i\lambda}\sigma_{i+}e^{i(\omega_0-\omega_{\lambda})t}- a^\dag_{i\lambda}\sigma_{i-}e^{-i(\omega_0-\omega_{\lambda})t})\nonumber \nonumber \\ &+ \sum_{j\neq i}J\sigma_{i}^+\sigma_{j}^-,
\end{align}
where we have an additional index on the annihilation and creation operators for the bosonic reservoir to denote each of the independent reservoirs associated to each atom.  This leads to the Laplace transform of $\Tilde{c}_i(s)$ 
\begin{align}
    \Tilde{c}_i(s) =& \frac{c_i(0)}{s-iJ+\Tilde{G}(s)} \nonumber \\
    &-\frac{iJc_+(0)}{(s+iJ(N-1)+\Tilde{G}(s))(s-iJ+\Tilde{G}(s))}.
\end{align}
In contrast to the previous section here we note that while there is a collective effect governed by the total polarisation $c_+$, we have also a decaying in the individual systems governed by $c_i(0)$. This behaviour is apparent from the fact that we can transform the first term into the single atom dynamics by way of taking $s\rightarrow s+iJ$ (simply introducing a phase parameter in the time domain) effectively shifting the transition energy of the atomic system by $iJ$. Furthermore, we note that the denominator for the term associated with the total polarisation no longer has a $N\Tilde{G}(s)$ term, showing that superradiant effects no longer play a role as the effective coupling strength no longer scales with number of atoms. If we utilise the Final Value Theorem as before we can see that, even with removal of the phase parameter $e^{iJt}$, the amplitudes vanish in steady-state conditions due to the non-zero nature of the Laplace transform of the memory kernel. 
\begin{equation}
    \lim_{s\rightarrow\infty} s\Tilde{c}_i(s) = 0.
\end{equation}
As such we no longer observe the steady state character we saw in the previous section. We can also note that the collective dynamics is a convolution of two systems undergoing decay into a reservoir as we now have multiple pathways for the excitation to leave the system and if the excitation enters another atom's reservoir it is now much more difficult for it to be retrieved by the other atoms, whereas previously all of the atoms may take the excitation from the same unique reservoir, now only one atom may do so, that corresponding to the reservoir the excitation is in. 
\subsection{Nearest Neighbour Coupling}
Similarly to the previous section we now consider the system wherein the atoms can only couple to adjacent atoms in a line. For this system the appropriate interaction Hamiltonian is given by
\begin{align}
    \Tilde{H}_I &= i\sum_{i,\lambda}g_{i\lambda}(a_{i\lambda}\sigma^+_i e^{i(\omega_0 -\omega_{i\lambda})t} - a^\dag_{i\lambda}\sigma^-_i e^{-i(\omega_0 -\omega_{i\lambda}t})\nonumber \\&+ \sum_{j=1}^{N-1}J(\sigma_j^+\sigma_{j+1}^- + \sigma_j^-\sigma_{j+1}^+).
\end{align}
Similar to the previous approach, we can evaluate the Laplace transform of the excited state amplitudes for the central atoms in the chain i.e for $\Tilde{c_i}, i\neq 1,N$ 
\begin{align}
    \Tilde{c}_i(s) =& \frac{c_i(0)(s+\Tilde{G}(s))-iJ(c_{i-1}(0)+c_{i+1}(0))}{(s+\Tilde{G}(s))^2+2J^2}\nonumber\\ &+\frac{J^2(c_{i-2}(0)+c_{i+2}(0))}{(s+\Tilde{G}(s))((s+\Tilde{G}(s))^2+2J^2)},
\end{align}
for $i\neq1,N$ this leads to the Laplace solutions for the first and final atom in the chain 
\begin{align}
\Tilde{c}_{1,N} = \frac{c_{1,N}(0)}{s+\Tilde{G}(s)} - \frac{iJ\Tilde{c}_i(s)}{s+\Tilde{G}(s)} .
\end{align}
Note that all terms contain the Laplace transform of the memory kernel $\Tilde{G}$ in the denominator, as such we expect them each to decay away into the reservoir and there isn't any observable shielding of these amplitudes. Clearly, in order to preserve non-vanishing steady-state excited state amplitudes, it is beneficial that the atomic systems interact with the same reservoir.

\section{Photonic Crystal Reservoir}
In this section we consider a specific model for the bosonic reservoir, that of the electromagnetic field around the photonic band gap of a photonic crystal. The photonic crystal is well suited for the model proposed as due to its screening of modes of the electromagnetic fields the low temperature approximation is well justified. Additionally, as to photonic crystals of interest have band gap frequencies in the optical spectral range,  the rotating wave approximation is also well justified Furthermore, experimental setups~\cite{KimbleOPhCr,SinglePhotonQD} have already been developed for studying atomic systems embedded within photonic crystals and as such we can compare our results to those found within experiments. The photonic crystal also has the additional useful quality that as we increase the dephasing value $\Delta$ (the difference between the atomic transition energy $\omega_0$ and the photonic band edge $\omega_I$) to large positive values we effectively see an unstructured vacuum and for large negatives values we see little of the electromagnetic field modes making the atoms act as free (no reservoir) coupled atoms. Thus, by varying the dephasing we can inspect a variety of different environment configurations and regimes to properly test our models. The model for the photonic crystal we have chosen is that of the isotropic one-dimensional photonic crystal band edge\cite{QEDNearPBG,John1994}. Such a system can generated in the form of a photonic crystal waveguide~\cite{PCWaveguide}. Such a model is interesting because it has a divergence in the local density of states of the electromagnetic field modes around the band edge frequency $\omega_C$ ($\rho(\omega)\propto\Theta(\omega-\omega_C)(\omega-\omega_C)^{-\frac{1}{2}}$) generating strong atom-photon coupling and leading to localisation of photons around atoms. 

\subsection{Fully Symmetric Coupling}
In order to perform our analysis we need only derive the memory kernel for the isotropic band gap model and then perform the inverse Laplace transforms required. For the memory kernel of the reservoir we have~\cite{SingleAtomSwitch}:
\begin{equation}
   \Tilde{G}(s) = \beta^{3/2} e^{-i\pi/4}(s-i\Delta)^{-\frac{1}{2}} ,
   \label{eqn:PCMemoryKernel}
\end{equation}
with the dephasing $\Delta$ defined by $\Delta = \omega_0 - \omega_I$ , where $\omega_I$ is the band edge frequency associated with the photonic band gap and $\beta$ is the coupling strength of the system-reservoir interaction which provides a canonical timescale for the dynamics. Substituting the photonic crystal memory kernel into Eq.~\ref{eqn:FCcis} yields for the excited state populations
\begin{align}
    \Tilde{c}_i(s) &= \frac{c_i(0)}{s-iJ} - \\
&\frac{c_+(0)(\beta^{3/2} e^{-i\pi/4}(s-i\Delta)^{-\frac{1}{2}} +iJ)}{(s-iJ)(s+iJ(N-1)+N\beta^{3/2} e^{-i\pi/4}(s-i\Delta)^{-\frac{1}{2}}) }.
\end{align}

Then by means of inverse Laplace transform we determine that the full time dynamics are given by
\begin{align}
    c_i(\tau) &= c_i(0)e^{iJ'\tau} - \\
    &c_+(0)e^{i\Delta' \tau}\sum_{i=1}^5 a_ix_ie^{x_i^2\tau}(1+r_i-r_i \text{Erfc}(\sqrt{x_i^2\tau})), 
    \label{eqn:invLTFC}
\end{align}
where $x_{1,2} = \pm e^{i\pi/4}\sqrt{J'-\Delta'}$ with $J'= J/\beta$, $\Delta'= \Delta'/\beta$,$\tau = t\beta$ being rescaled by the canonical time $\beta$ and are dimensionless,
\begin{align}
    x_3 &= (A_+ + A_-)e^{i\pi/4},\nonumber  \\
    x_4 &= (A_+e^{-i\pi/6} - A_-e^{i\pi/6})e^{-i\pi/4},\nonumber \\
    x_5 &= (A_+e^{i\pi/6} - A_-e^{-i\pi/6})e^{3i\pi/4},\nonumber \\
    A_\pm &= N^{1/3}\left[\frac{1}{2} \pm \frac{1}{2}\left(1 + \frac{4\Tilde{\kappa}^3}{27N^2}\ \right)^\frac{1}{2} \right]^\frac{1}{3},\nonumber \\
    \Tilde{\kappa} &= (\Delta' + J'(N-1)),\nonumber \\
    a_k &= (e^{-i\pi/4}+ iJ'x_i)\prod^N_{j\neq i} \frac{1}{x_i-x_j}, \nonumber \\
    r_k &= \text{csgn}(x_k),
    \label{eqn:PCFC}
\end{align}
with $\text{csgn}(x_i)$ being the complex sign function.

For convenience moving forward we will drop the superscript denoting the dimensionless dipole-dipole coupling and dephasing values. 
From Eq.~\ref{eqn:PCFC}  we note that if we remove the dipole-dipole coupling by taking $J\rightarrow 0$ then we note that $A_\pm \propto N^{\frac{1}{3}}$ this implies that $x_i^2 \propto N^{\frac{2}{3}}$. These values are associated with exponential decay in Eq.~\ref{eqn:invLTFC},  hence we conclude that the decay rate $\gamma \propto N^{\frac{2}{3}}$ reproduces the result found in~\cite{SuperRadiancePBG} for uncoupled atomic systems inside a 1D photonic crystal, thus demonstrating superradiant behaviour even in the single-excitation regime. Conversely, if we consider non-zero values for the dipole-dipole coupling $J$ then for large numbers of atomic systems $N$ or small values of the dephasing $\Delta$,  $A_\pm \propto N^{\frac{1}{2}}$ leading to $x_i^2 \propto N$ effectively increasing the rate of superradiance by introducing dipole-dipole coupling. Remarkably this relationship has been observed for trapped atoms inside of 1D photonic wave-guides where the decay rate was found to be proportional to the number of atoms $N$~\cite{KimbleSuperradiance} hence validating our approach.

\begin{figure}
    \centering
    \includegraphics[width=8.6cm]{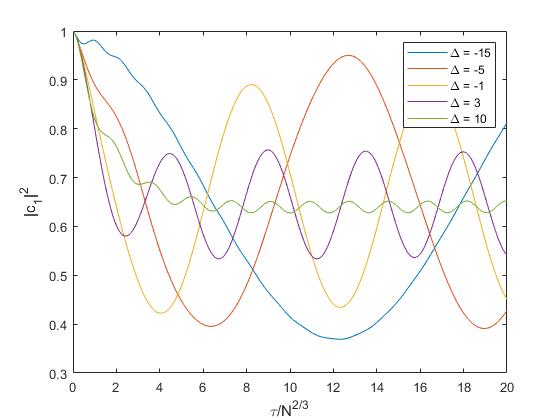}
    \caption{Population dynamics for the excited state of a single atom coupled to 4 other atoms, with the atom initially excited ($c_1(0)=1$), inside of a photonic crystal reservoir, with varying values of the dephasing parameter $\Delta = \omega_0 -\omega_C$ between the atomic transition energy and the photonic crystal band edge, the dipole-dipole coupling $J=0.1$.}
    \label{fig:FCatom1n5}
\end{figure}

\begin{figure}
    \centering
    \includegraphics[width=8.6cm]{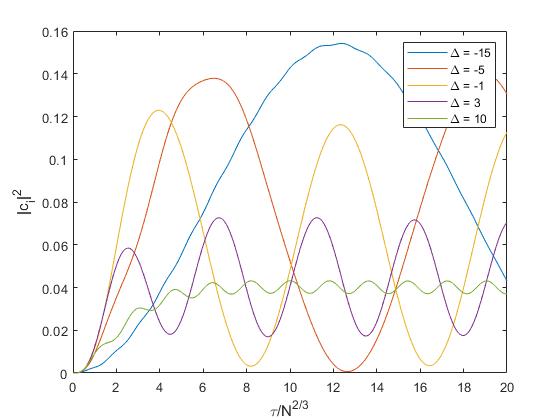}
    \caption{Population dynamics for the excited state of a single atom coupled to 4 other atoms, with another atom initially excited ($c_1(0)=1, i\neq 1$), inside of a photonic crystal reservoir, with varying values of the dephasing parameter $\Delta = \omega_0 -\omega_C$ between the atomic transition energy and the photonic crystal band edge, the dipole-dipole coupling $J=0.1$.}
    \label{fig:FCatomin5}
\end{figure}

\begin{figure}
    \centering
    \includegraphics[width=8.6cm]{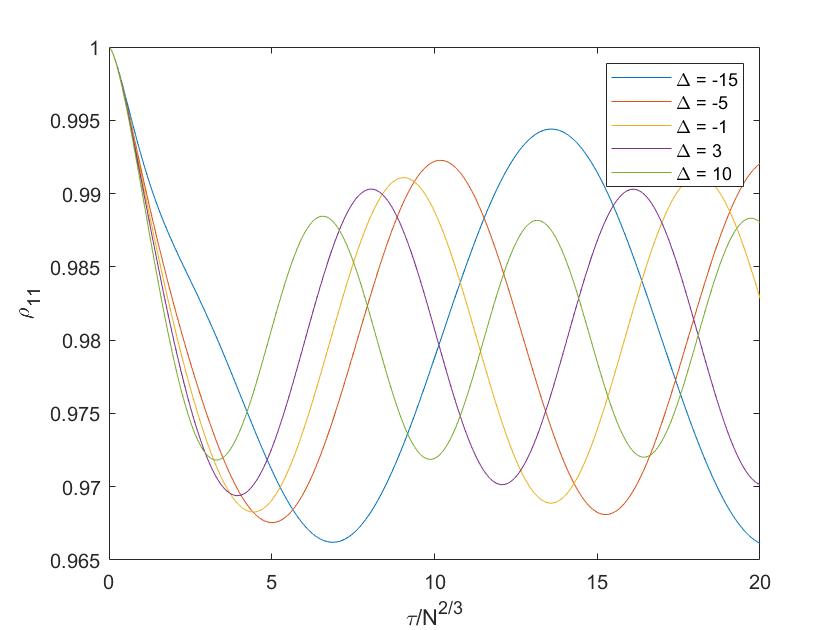}
    \caption{Population dynamics for the excited state of a single atom coupled to 99 other atoms, with the atom initially excited ($c_1(0)=1$), inside of a photonic crystal reservoir, with varying values of the dephasing parameter $\Delta = \omega_0 -\omega_C$ between the atomic transition energy and the photonic crystal band edge, the dipole-dipole coupling $J=0.1$.}
    \label{fig:FCatom1n100}
\end{figure}

From \Cref{fig:FCatom1n5,fig:FCatomin5,fig:FCatom1n100} (here atom 1 is initially excited), we note the predicted steady state values (minus oscillations) of $1/N^2$ are present. As we increase the value of the dephasing $\Delta$, we move closer to a vacuum state so would traditionally expect the atom to de-excite into the the ground state. Additionally, due to the nature of the photonic crystal we have non-zero steady states associated with the total polarisation (acting as the single atom system from~\cite{SingleAtomSwitch}) of the system. As such, we still have oscillations as the excitation is passed between the different atoms. these amplitudes become damped as we increase the dephasing value, due to a monotonic reduction in the total polarisation in $\Delta$.  We also note that in Fig.~\ref{fig:FCatom1n100} as the number of atoms increases the atoms move closer and closer in phase as $\Delta +J(N-1) \to JN$.
\begin{figure}
    \centering
    \includegraphics[width=8.6cm]{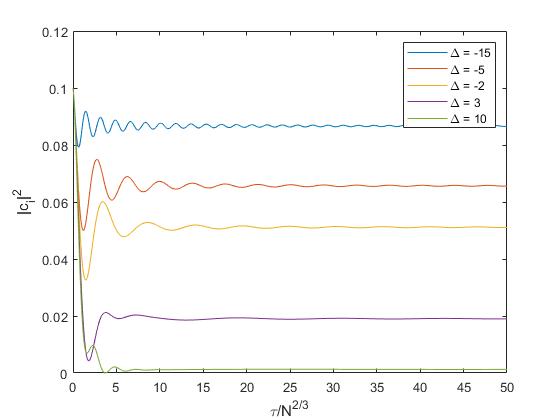}
    \caption{Population dynamics for the excited state of a single atom coupled to 9 other atoms, with all atoms symmetrically initially prepared ($c_i(0)=1/\sqrt{10}$), inside of a photonic crystal reservoir, with varying values of the dephasing parameter $\Delta = \omega_0 -\omega_C$ between the atomic transition energy and the photonic crystal band edge, the dipole-dipole coupling $J=0.1$.}
    \label{fig:FCatomin10FS}
\end{figure}

\begin{figure}
    \centering
    \includegraphics[width=8.6cm]{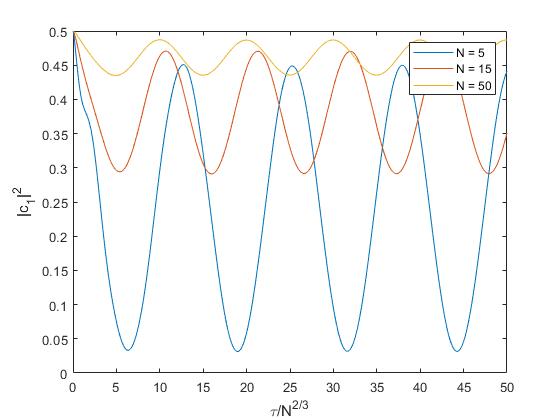}
    \caption{Dynamics of the Bell state $\Psi_+$ for varying numbers of atoms $N$ with value of the dephasing parameter $\Delta = \omega_0 -\omega_C = -5$ between the atomic transition energy and the photonic crystal band edge. }
    \label{fig:Psi+}
\end{figure}

Fig.~\ref{fig:FCatomin10FS} shows that when the excited state amplitudes for all atoms are initially prepared in a symmetrical excited state ($c_i(0)=1/\sqrt{10}$ for all $i$),  we return to the single atom ($N=1,J=0$) excited state amplitude's dynamics multiplied by $1/N$ with a Lamb shifted transition energy given by
\begin{equation}
\Delta \to \frac{\Delta+J(N-1)}{N^{2/3}},
\end{equation}
as predicted.
 As such we no longer see any transfer of excitations between respective atoms as each atom loses as much polarisation as it gains from the other atoms in the system reaching an equilibrium. Additionally, as we increase the number of atoms $N$ we effectively shift the atoms outside of the band gap as
 \begin{equation}
    N>>\Delta/J, \kappa = \frac{\Delta+J(N-1)}{N^{2/3}}\to JN^{1/3}. 
 \end{equation}
 Next we analyse the coherences between the energy eigenstates of the atomic systems. This is done by  exploring the dynamics of the canonical Bell states. Typically, the environment causes decoherence in quantum systems. If we could identify environments that better preserve the coherences in atomic systems we may be able to engineer more robust quantum computing systems.
 The pair of Bell states accessible in this regime are 
\begin{align}
\Psi_+ &= \frac{1}{\sqrt{2}}( \ket{0}\ket{1} + \ket{1}\ket{0}), \nonumber\\
\Psi_- &= \frac{1}{\sqrt{2}}( \ket{0}\ket{1} - \ket{1}\ket{0}) .
\end{align}
If we consider a system of N atoms and we consider only two such atoms, we may choose to prepare the system in such a way that the two atoms have initial amplitudes $c_1(0)=\pm \frac{1}{\sqrt{2}} $ and $c_2(0)= \frac{1}{\sqrt{2}}$ such that the atoms are initialised in either of the two Bell states tensor product with the ground states of the $N-2$ other atoms and the reservoir. From Eqn.\ref{eqn:invLTFC}  we can see that for the $\Psi_-$ state - as $c_+(0)=c_1(0)+c_2(0)=0$ - this system is decoupled from its environment as is well conserved. However, for the initialisation such that the two atoms are in the $\Psi_+$ state $c_+(0)=\sqrt{2}$, and thus we expect non-trivial reservoir induced dynamics in the states evolution. However, as previously discussed, by increasing the number of atoms in the system we can reduce the effects of the reservoir on individual atoms. From Fig.~\ref{fig:Psi+} note  that as we increase the number of atoms in the system $N$ the oscillations away from the initial condition of the system are reduced, suggesting that increased auxiliary atoms may provide better conservation of the Bell state of the two atom system.
\subsection{Nearest Neighbour  Coupling}
\begin{figure}
    \centering
    \includegraphics[width=8.6cm]{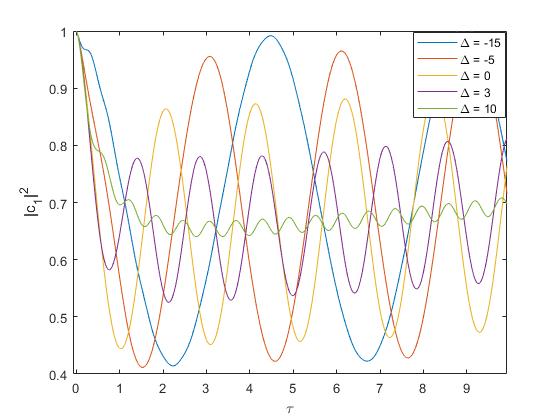}
    \caption{Population dynamics for the excited state of the first atom in a chain of 5 atoms coupled to their nearest neighbour, with the first atom initially excited ($c_1(0)=1$), inside of a photonic crystal reservoir, with varying values of the dephasing parameter $\Delta = \omega_0 -\omega_C$ between the atomic transition energy and the photonic crystal band edge, the dipole-dipole coupling $J=0.1$.}
    \label{fig:NNa1n5}
\end{figure}

\begin{figure}
    \centering
    \includegraphics[width=8.6cm]{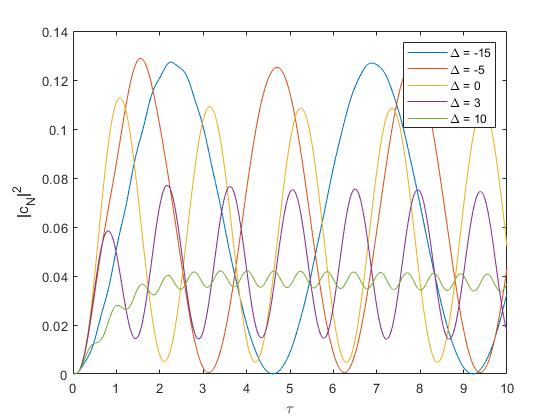}
    \caption{Population dynamics for the excited state of the final atom in a chain of 5 atoms coupled to their nearest neighbour, with the first atom in the chain initially excited ($c_1(0)=1$), inside of a photonic crystal reservoir, with varying values of the dephasing parameter $\Delta = \omega_0 -\omega_C$ between the atomic transition energy and the photonic crystal band edge, the dipole-dipole coupling $J=0.1$.}
    \label{fig:NNaNn5}
\end{figure}

We now consider the nearest neighbour coupling of a chain of atoms embedded within a photonic crystal. Similar to the previous approach,  we substitute the 1D photonic crystal memory kernel from Eq.~\ref{eqn:PCMemoryKernel} into Eq.~\ref{eqn:NNcN} and perform the inverse Laplace transform which yields
\begin{equation}
    c_N(\tau) = -e^{i\Delta \tau}\sum_{i=1}^N a_ix_ie^{x_i^2\tau}(1+r_i-r_i \text{Erfc}(\sqrt{x_i^2\tau})) ,
\end{equation}
where $x_i$ are the 5 complex roots of the equation 
\begin{align}
    x^5 + x^3(2i\Delta +2iJ) + x^2Ne^{-i\pi/4} - x(\Delta^2 -2J\Delta)\nonumber \\+2iJe^{-i\pi/4} + i\Delta Ne^{-i\pi/4} = 0,
\end{align}
$r_i = \text{csgn}(x_i)$ and 
\begin{equation}
    a_i = e^{-i\pi/4}\prod^N_{j\neq i} \frac{1}{x_i-x_j}.
\end{equation}
For the dynamics of the first atom in the chains excited state amplitude we have
\begin{equation}
    c_1(\tau) = c_1(0) +c_N(\tau).
\end{equation}
Since the values of $x_i$ are the complex roots of a 5$^\textrm{th}$ order polynomial equation, we opt for a numerical solution. Figs.~\ref{fig:NNa1n5} and~\ref{fig:NNaNn5} depict the dynamics of the Bell state and of the first atom in a 5-atom chain, and we note that the results for this nearest neighbour configuration are very similar to the dynamics of the fully coupled system. Clearly, it becomes advantageous to explore as much as possible the fully coupled system, which has an analytical solutions and extrapolate the results to the nearest neighbour coupled system. This behaviour also suggest that the dynamics of the atomic system is being more significantly driven by the reservoir than the dipole-dipole coupling.

\section{Conclusion}
In this article we have analysed the many-body dissipative systems coupled to non-Markovian reservoirs and explored  novel features of such systems. We have focused on the the steady state character of the atomic dynamics that is determined by the number of atoms within the system this effect and is shown to be independent of the type of reservoir the atoms are interacting with as well as the  dipole-dipole coupling strengths. 
We further expanded our analysis to the case of nearest neighbour coupling between atoms to model an atomic chain and identified a similar steady state,  this time non-trivially determined by the  dipole-dipole coupling. In contrast to coupling to independent reservoirs, wherein no such steady state can be achieved,  we identify multiple channels for the system to dissipate into without retrieval. Finally, we have explored the isotropic photonic band gap model and have shown that it is possible to observe both the late time dynamics predicted as well as superradiant behaviour. We have shown that the superradiance can be enhanced by dipole-dipole coupling between the atomic systems and have both reproduced previous theoretical results, in the absence of dipole-dipole coupling, as well as found agreement with experimental results when reintroducing the dipole-dipole coupling~\cite{KimbleSuperradiance} suggesting the necessity of this interaction in understanding the physical system. Our results demonstrates the viability of the approach introduced and suggests that such models may have physical constructions that could be of practical interest as they can effectively manipulate the localisation of excitations across atoms simply by virtue of having more atoms supporting them ~\cite{ManyAtomsInPCW}. Furthermore, we have shown that in this paradigm the single excitation Bell states $\Psi_\pm$ can be well preserved and may prove useful for quantum computational systems. 

\begin{acknowledgments}
This work was supported by the Leverhulme Quantum Biology Doctoral Training Centre at the University of Surrey were funded by a Leverhulme Trust training centre grant number DS-2017-079, and the EPSRC (United Kingdom) Strategic Equipment Grant No. EP/L02263X/1 (EP/M008576/1) and EPSRC (United Kingdom) Grant EP/M027791/1 awards to M.F. We acknowledge helpful discussions with the members of the Leverhulme Quantum Biology Doctoral Training Centre.
\end{acknowledgments}

\appendix

\section{Single Reservoir}
\subsection{Fully Symmetric Coupling}
We start with the (RWA) Hamiltonian 
\begin{align}
    H =& \sum_{i=1}^N\omega_0\sigma_i^+\sigma_i^- + \sum_\lambda \omega_\lambda a^\dag_\lambda a_\lambda \nonumber\\&+ i \sum_{i,\lambda}g_\lambda(a_\lambda \sigma^+_i -a^\dag_\lambda \sigma^-_i ) + \sum_{i\neq j}J\sigma^+_i\sigma_j^-,
\end{align}
where $\sigma_i^+$and $\sigma_i^-$, are the excitation and de-excitation operators for the $i$th atomic system respectively, $a_{\lambda}$ and $a^{\dagger}_{\lambda}$ are the bosonic field annihilation and creation operators, $\omega_0$ and $\omega_{\lambda}$ are the atomic transition and the $\lambda$-boson mode frequencies, and $g_\lambda$ is the coupling strength of the atomic system and the $\lambda$-boson mode.

By changing to the interacting picture we can consider the interaction Hamiltonian 
\begin{align}
    \Tilde{H}_I =& i\sum_{i,\lambda}g_\lambda(a_\lambda \sigma^+_i e^{i(\omega_0-\omega_\lambda)t} -a^\dag_\lambda \sigma^-_i e^{-i(\omega_0-\omega_\lambda)t} )\nonumber \\& + \sum_{j\neq k}J\sigma^+_j\sigma_k^-.
\end{align}
Such a Hamiltonian is convenient as it conserves the excitation number of any wavefunction. As such if we consider the single excitation wavefunction given by 
\begin{equation}
    \phi(0) = c_0\psi_0+\sum_i^N c_i(0)\psi_i +\sum_\lambda c_\lambda(0) \psi_\lambda,
\end{equation}
it's time evolution is given by 
\begin{equation}
    \phi(t) = c_0\psi_0 +\sum_i^N c_i(t)\psi_i +\sum_\lambda c_\lambda(t) \psi_\lambda,
\end{equation}
where $\psi_i = \ket{i}_A\ket{0}_B$ the ith atom is in its excited state and $\psi_\lambda = \ket{0}_A\ket{\lambda}_B$ the atoms are all in their ground state and the bosonic system has its $\lambda$ mode excited. By simply plugging this into the Schr{\"o}dinger equation and determining the Heisenberg equations of motion for the state amplitudes we get 
\begin{equation}
    \Tilde{H}_I \phi_i = -i\sum_\lambda g_\lambda \psi_\lambda e^{-i(\omega_0-\omega_\lambda)t} + \sum_j^N J \psi_j (1- \delta_{ij}),
\end{equation}

\begin{equation}
    \Tilde{H}_I \phi_\lambda = ig_\lambda e^{i(\omega_0-\omega_\lambda)t} \sum_i \psi_i.
\end{equation}

By introducing the parameter 
\begin{equation}
    c_+(t) = \sum_i c_i(t)  , 
\end{equation}
we have 
\begin{align}
    \Dot{c}_i =& -iJ(c_+-c_i) + \sum_\lambda c_\lambda g_\lambda e^{i(\omega_0-\omega_\lambda)t},\nonumber \\
    \Dot{c}_+ =& -iJ(N-1)c_+ + N \sum_\lambda c_\lambda g_\lambda e^{i(\omega_0-\omega_\lambda)t} ,\nonumber \\
    \Dot{c}_\lambda =& -g_\lambda e^{-i(\omega_0-\omega_\lambda)t} c_+.
\end{align}
Assuming the bosonic field is initially in its vacuum configuration ($c_\lambda(0) = 0$) we can formally integrate up the last equation to get 
\begin{equation}
    c_\lambda(t) = -\int^t_0 g_\lambda e^{-(\omega_0-\omega_\lambda)t_1}c_+(t_1)dt_1,
\end{equation}
and by introducing the memory kernel $G(t) = \sum_\lambda g_\lambda^2 e^{i(\omega_0-\omega_\lambda)t}$ we can rewrite the above equations as 

\begin{align}
    \Dot{c}_i =& -iJ(c_+-c_i) - \int^t_0G(t-t_1)c_+(t_1)dt_1,\nonumber \\
    \Dot{c}_+ =& -iJ(N-1)c_+ - N\int^t_0G(t-t_1)c_+(t_1)dt_1.    
\end{align}
It is notable that our memory kernel $G(t)$ is related to the spectral density $J(\omega)= \sum_\lambda g_\lambda^2 \delta(\omega-\omega_\lambda)$ associated to the reservoir by the relation
\begin{equation}
G(t) = \int d\omega J(\omega )e^{i(\omega_0-\omega)t}.
\end{equation}
In order to solve these equations we utilise the Laplace transform. Transforming the differential equations into algebraic ones.

\begin{align}
    s\Tilde{c}_i(s) - c_i(0) &= -iJ(\Tilde{c}_+(s) -\Tilde{c}_i(s)) - \Tilde{c}_+(s)\Tilde{G}_1\nonumber \\
    s\Tilde{c}_+(s) - c_+(0) &= -iJ(N-1)\Tilde{c}_+(s) - N\Tilde{c}_+(s)\Tilde{G}_1.
\end{align}
Where we have used that $\Tilde{f}(s)=\mathcal{L}\{f(t)\}$. We can see that we can solve for $\Tilde{c}_+(s)$ getting

\begin{equation}
    \Tilde{c}_+(s) = \frac{c_+(0)}{s+iJ(N-1)+N\Tilde{G}}
\end{equation}
From this we can solve for the single atom amplitude Laplace solution 
\begin{equation}
    \Tilde{c}_i(s) = \frac{c_i(0)}{s-iJ} - \frac{c_+(0)(\Tilde{G}+iJ)}{(s-iJ)(s+iJ(N-1)+N\Tilde{G})}.
\end{equation}
If we consider the form of $\Tilde{G}(s)$ with respect to the spectral density we can see that

\begin{equation}
\Tilde{G}(s) =\int d\omega \frac{J(\omega)}{s-i(\omega_0-\omega)}.
\end{equation}

\subsection{Nearest Neighbour coupling}
We start with the interaction Hamiltonian for the nearest neighbour coupling,  wherein the atoms only couple to those either side of them, such an interaction Hamiltonian is of the form

\begin{align}
    \Tilde{H}_I =& i\sum_{i,\lambda}g_\lambda(a_\lambda\sigma^+_i e^{i(\omega_0 -\omega_\lambda)t} - a^\dag_\lambda\sigma^-_i e^{-i(\omega_0 -\omega_\lambda)t})\nonumber\\& + \sum_{j=1}^{N-1}J(\sigma_j^+\sigma_{j+1}^- + \sigma_j^-\sigma_{j+1}^+)
\end{align}
by performing a similar process as for the symmetric case we find that the Heisenberg equations of motion are
\begin{equation}
    \Dot{c}_\lambda = -g_\lambda e^{-i(\omega_0-\omega_\lambda)t} c_+
\end{equation}
\begin{align}
    \Dot{c}_i = -iJ((1-\delta_{N,i})c_{i+1}+(1-\delta_{1,i})c_{i-1}) \nonumber\\- \int\limits^t_0 dt_1 c_+(t_1)G(t-t_1)
\end{align}
we then integrate up the state amplitudes associated with environment excitations to determine the total polarisation parameter
\begin{equation}
    \Dot{c}_+ =  -N\int\limits^t_0 dt_1 c_+(t_1)G(t-t_1) - iJ(2c_+- c_0 -c_N)
\end{equation}
substituting $c_i$ into itself to 2nd order in the dipole-dipole coupling $J$ yields 
\begin{equation}
    \Tilde{c}_+(s) = \frac{s^2\Tilde{c}_+(0) + iJs(c_1(0)+c_N(0))+J^2(c_2(0)+c_{N-1}(0))}{s^3+(N\Tilde{G}+2iJ)s^2-2iJ\Tilde{G}s+2J^2\Tilde{G}(s)}
\end{equation}
leading to the equation for the atom 1 excited state amplitude
\begin{align}
    \Tilde{c}_1(s) =& \frac{s^2c_1(0)-iJsc_2(0)-J^2c_3(0)}{(s^2+J^2)s}\nonumber\\+&\frac{\Tilde{c}_+(s)\Tilde{G}(s)(J^2+iJs-s^2)}{(s^2+J^2)s}
\end{align}
and for atom N
\begin{align}
    \Tilde{c}_N(s) = & \frac{s^2c_N(0)-iJsc_{N-1}(0)-J^2c_{N-2}(0)}{(s^2+J^2)s}\nonumber\\ &+\frac{\Tilde{c}_+(s)\Tilde{G}(s)(J^2+iJs-s^2)}{(s^2+J^2)s}.
\end{align}

\subsection{Nearest Neighbour differential ends}
We start with the interaction Hamiltonian for the nearest neighbour coupling,  wherein the atoms only couple to those either side of them, however, the first and last atom in the chain have a transition energy $\omega_0+2\delta$ and $\omega_0$ respectively, with the intermediate atoms having transition energy $\omega_0+\delta$. 
The interaction Hamiltonian is of the form
\begin{align}
    \Tilde{H}_I =& J(\sigma_1^+\sigma_2^-e^{i\delta t} + \sigma_{N-1}^+\sigma^-_{N}e^{i\delta t} +  \sum^{N-2}_{n=2}\sigma^+_n\sigma^-_{n+1}) \nonumber  \nonumber \\+    &i \sum_\lambda g_\lambda(\sigma^+_1 a_\lambda e^{i(\omega +2\delta - \omega_\lambda)t}  + \sigma^+_N a_\lambda e^{i(\omega - \omega_\lambda)t} \nonumber\\& + \sum_{n=2}^{N-2}\sigma^+_n a_\lambda e^{i(\omega +\delta - \omega_\lambda)t}) +\text{h.c}
\end{align}
where h.c represents the Hermitian conjugate. The associated Heisenberg equations of motion are given by 
\begin{align}
    \Dot{c}_1 &= -ic_2Je^{i\delta t} -\sum_\lambda c_\lambda g_\lambda e^{i(\omega +2\delta -\omega_\lambda)t}\nonumber \\
    \Dot{c}_i &= -iJ(c_{i-1} +c_{i+1}) -\sum_\lambda c_\lambda g_\lambda e^{i(\omega +\delta -\omega_\lambda)t}\nonumber \\
        \Dot{c}_N &= -ic_{N-1}Je^{-i\delta t} -\sum_\lambda c_\lambda g_\lambda e^{i(\omega -\omega_\lambda)t}\nonumber \\
        \Dot{c}_\lambda &= -g_\lambda(c_1e^{-i(\omega +2\delta-\omega_\lambda)t } + \sum_n c_ne^{-i(\omega +\delta-\omega_\lambda)t } \nonumber\\&+c_Ne^{-i(\omega-\omega_\lambda)t }
        \nonumber \\
        c_\lambda &= -g_\lambda\int^t_0 dt_1 e^{-i(\omega - \omega_\lambda)t_1}(c_1(t_1) e^{-2i\delta t} \nonumber\\&+ \sum_{n=2}^{N-2} c_n(t_1)e^{-i\delta t} + c_N(t_1) ) 
\end{align}
We define $c_m(t) = \sum_{n=2}^{N-2}c_n$ and moving forward drop order $J^2$ and higher terms. We also introduce the notation
\begin{equation}
    G_j(t) = G(t)e^{i(j-1)\delta t}
\end{equation}

and noting that $\Tilde{G}_j(s+i\delta) = \Tilde{G}_{j-1}(s)$. We may now consider the Laplace transform solutions 
\begin{align}
    \Dot{c}_1 &= -iJc_2e^{i\delta t}-\int^t_0 dt_1 G_3(t-t_1)(c_1(t_1) \nonumber\\&+ \sum_{n=2}^{N-2}c_i(t_1) e^{i\delta t_1} +c_N(t_1)e^{2\delta it}
    \nonumber \\
    s\Tilde{c}_1(s) &-c_1(0) = -iJ\Tilde{c}_2(s-i\delta) -\Tilde{G}_3(s)(\Tilde{c}_1(s) \nonumber\\&+  \Tilde{c}_m(s-i\delta) + \Tilde{c}_N(s-2i\delta))
    \nonumber \\
    s\Tilde{c}_i(s) &= -iJ(\Tilde{c}_{i-1}(s) + \Tilde{c}_{i+1}(s)) - \Tilde{G}_2(s)(\Tilde{c}_1(s+i\delta) \nonumber\\&+ \Tilde{c}_m(s) + \Tilde{c}_N(s-i\delta))
\end{align}
and for convenience we introduce a set of equations to simplify the derivation
\begin{align}
    p(s) &= \Tilde{G}_2(s)(N-2)+s+2iJ(1+\frac{\Tilde{G}_2(s)}{s})\nonumber \\
    q(s) &= iJ(2\frac{\Tilde{G}_2(s)}{s}+1) +(N-2)\Tilde{G}_2(s)\nonumber \\
    \Tilde{c}_m(s) &= -\frac{p(s)}{q(s)}(\Tilde{c}_1(s+i\delta) + \Tilde{c}_N(s-i\delta))\nonumber \\
    M(s) &= \Tilde{G}_1(s)(\frac{iJ}{s+i\delta}-1)(1-\frac{p(s+i\delta)}{q(s+i\delta)})\nonumber \\
    \Tilde{c}_N(s) &= \frac{M(s)c_1(0)}{s^2 +2i\delta s - 2(s+i\delta) M(s)}\nonumber \\
    &= \frac{\Tilde{G}_1(s)(iJ-s-i\delta)(s+i\delta+iJ)}{(s+i\delta)(q(s+i\delta)+2\Tilde{G}_1(s))} 
\end{align}
if we assume small value for $\delta$ such that the rotating wave approximation is valid we have (i.e dropping $\delta^2$ terms)
\begin{equation}
    \Tilde{c}_N(s) = \frac{-\Tilde{G}_1(s)}{(s+i\delta)^2 + N\Tilde{G}_1(s)(s+i\delta) +2iJ(\Tilde{G}_1(s)+ (s+i\delta))}.
\end{equation}
This is just a phase shifted version of the same transition energy case we study previously. 

\section{Independent Reservoirs}
\subsection{Fully Symmetric Coupling}
We now consider the interaction Hamiltonian for independent reservoirs which requires us to index over the various atoms reservoir operators.
The interaction Hamiltonian then becomes
\begin{align}
    \Tilde{H}_I =& i\sum_{i\lambda}g_{\lambda}(a_{i\lambda}\sigma_{i}^+e^{i(\omega_0-\omega_{\lambda})t}- a^\dag_{i\lambda}\sigma_{i}^-e^{-i(\omega_0-\omega_{\lambda})t})\nonumber \nonumber \\ &+ \sum_{j\neq i}J\sigma_{i}^+\sigma_{j}^-
\end{align}
where we have assumed each reservoir to have the same spectrum of modes such that they all couple with strength $g_\lambda$ and $a_{i\lambda}$ ($a^\dag_{i\lambda}$) are the annhilation (creation) operators for the $\lambda$ mode of the $i$th atom's reservoir.
Now we need to consider the wavefunction taking a different form as we must index the reservoir to which excitations can go, as such we have our total wavefunction
\begin{equation}
    \phi(t) = c_0\psi_0+\sum_i^N c_i(t)\psi_i +\sum_{i\lambda} c_{i\lambda}(t) \psi_{i\lambda}
\end{equation}
where $\psi_{i\lambda}$ denotes the excitation being in the $\lambda$ mode of the $i$th atom's reservoir. The Heisenberg equations of motion are then

\begin{align}
    \Dot{c}_i =& -iJ(c_+-c_i) + \sum_\lambda c_{i\lambda} g_\lambda e^{i(\omega_0-\omega_\lambda)t}\nonumber \\
    \Dot{c}_+ =& -iJ(N-1)c_+ + \sum_{i\lambda} c_{i\lambda} g_\lambda e^{i(\omega_0-\omega_\lambda)t} \nonumber \\
    \Dot{c}_{i\lambda} =& -g_\lambda e^{-i(\omega_0-\omega_\lambda)t} c_i
\end{align}
formally integrating up $c_{i\lambda}$ with $c_{i\lambda}(0)=0$ and substituting back in 

\begin{align}
    \Dot{c}_i =& -iJ(c_+-c_i) - \int^t_0G(t-t_1)c_i(t_1)dt_1\nonumber \\
    \Dot{c}_+ =& -iJ(N-1)c_+ - \int^t_0G(t-t_1)c_+(t_1)dt_1 \nonumber \\    
\end{align}
then from this we can determine the Laplace transform solutions,
\begin{align}
    s\Tilde{c}_i(s) - c_i(0) &= -iJ(\Tilde{c}_+(s) -\Tilde{c}_i(s)) - \Tilde{c}_i(s)\Tilde{G}(s)\nonumber \\
    s\Tilde{c}_+(s) - c_+(0) &= -iJ(N-1)\Tilde{c}_+(s) - \Tilde{c}_+(s)\Tilde{G}(s).
\end{align}
which gives 
\begin{equation}
    \Tilde{c}_+(s) = \frac{c_+(0)}{s+iJ(N-1)+\Tilde{G}(s)}
\end{equation}
solving then for $\Tilde{c}_i(s)$ yields

\begin{equation}
    \Tilde{c}_i(s) = \frac{c_i(0)}{s-iJ+\Tilde{G}(s)} -\frac{iJc_+(0)}{(s+iJ(N-1)+\Tilde{G}(s))(s-iJ+\Tilde{G}(s))}.
\end{equation}

\bibliography{main}

\begin{thebibliography}{39}%
\makeatletter
\providecommand \@ifxundefined [1]{%
 \@ifx{#1\undefined}
}%
\providecommand \@ifnum [1]{%
 \ifnum #1\expandafter \@firstoftwo
 \else \expandafter \@secondoftwo
 \fi
}%
\providecommand \@ifx [1]{%
 \ifx #1\expandafter \@firstoftwo
 \else \expandafter \@secondoftwo
 \fi
}%
\providecommand \natexlab [1]{#1}%
\providecommand \enquote  [1]{``#1''}%
\providecommand \bibnamefont  [1]{#1}%
\providecommand \bibfnamefont [1]{#1}%
\providecommand \citenamefont [1]{#1}%
\providecommand \href@noop [0]{\@secondoftwo}%
\providecommand \href [0]{\begingroup \@sanitize@url \@href}%
\providecommand \@href[1]{\@@startlink{#1}\@@href}%
\providecommand \@@href[1]{\endgroup#1\@@endlink}%
\providecommand \@sanitize@url [0]{\catcode `\\12\catcode `\$12\catcode
  `\&12\catcode `\#12\catcode `\^12\catcode `\_12\catcode `\%12\relax}%
\providecommand \@@startlink[1]{}%
\providecommand \@@endlink[0]{}%
\providecommand \url  [0]{\begingroup\@sanitize@url \@url }%
\providecommand \@url [1]{\endgroup\@href {#1}{\urlprefix }}%
\providecommand \urlprefix  [0]{URL }%
\providecommand \Eprint [0]{\href }%
\providecommand \doibase [0]{https://doi.org/}%
\providecommand \selectlanguage [0]{\@gobble}%
\providecommand \bibinfo  [0]{\@secondoftwo}%
\providecommand \bibfield  [0]{\@secondoftwo}%
\providecommand \translation [1]{[#1]}%
\providecommand \BibitemOpen [0]{}%
\providecommand \bibitemStop [0]{}%
\providecommand \bibitemNoStop [0]{.\EOS\space}%
\providecommand \EOS [0]{\spacefactor3000\relax}%
\providecommand \BibitemShut  [1]{\csname bibitem#1\endcsname}%
\let\auto@bib@innerbib\@empty
\bibitem [{\citenamefont {Petta}\ \emph {et~al.}(2005)\citenamefont {Petta},
  \citenamefont {Johnson}, \citenamefont {Taylor}, \citenamefont {Laird},
  \citenamefont {Yacoby}, \citenamefont {Lukin}, \citenamefont {Marcus},
  \citenamefont {Hanson},\ and\ \citenamefont
  {Gossard}}]{CoherentSCQuantumDots}%
  \BibitemOpen
  \bibfield  {author} {\bibinfo {author} {\bibfnamefont {J.~R.}\ \bibnamefont
  {Petta}}, \bibinfo {author} {\bibfnamefont {A.~C.}\ \bibnamefont {Johnson}},
  \bibinfo {author} {\bibfnamefont {J.~M.}\ \bibnamefont {Taylor}}, \bibinfo
  {author} {\bibfnamefont {E.~A.}\ \bibnamefont {Laird}}, \bibinfo {author}
  {\bibfnamefont {A.}~\bibnamefont {Yacoby}}, \bibinfo {author} {\bibfnamefont
  {M.~D.}\ \bibnamefont {Lukin}}, \bibinfo {author} {\bibfnamefont {C.~M.}\
  \bibnamefont {Marcus}}, \bibinfo {author} {\bibfnamefont {M.~P.}\
  \bibnamefont {Hanson}},\ and\ \bibinfo {author} {\bibfnamefont {A.~C.}\
  \bibnamefont {Gossard}},\ }\href {https://doi.org/10.1126/science.1116955}
  {\bibfield  {journal} {\bibinfo  {journal} {Science}\ }\textbf {\bibinfo
  {volume} {309}},\ \bibinfo {pages} {2180} (\bibinfo {year}
  {2005})}\BibitemShut {NoStop}%
\bibitem [{\citenamefont {Yamamoto}\ \emph {et~al.}(2003)\citenamefont
  {Yamamoto}, \citenamefont {Pashkin}, \citenamefont {Astafiev}, \citenamefont
  {Nakamura},\ and\ \citenamefont {Tsai}}]{GateOperationQubits}%
  \BibitemOpen
  \bibfield  {author} {\bibinfo {author} {\bibfnamefont {T.}~\bibnamefont
  {Yamamoto}}, \bibinfo {author} {\bibfnamefont {Y.~A.}\ \bibnamefont
  {Pashkin}}, \bibinfo {author} {\bibfnamefont {O.}~\bibnamefont {Astafiev}},
  \bibinfo {author} {\bibfnamefont {Y.}~\bibnamefont {Nakamura}},\ and\
  \bibinfo {author} {\bibfnamefont {J.~S.}\ \bibnamefont {Tsai}},\ }\href
  {https://doi.org/10.1038/nature02015} {\bibfield  {journal} {\bibinfo
  {journal} {Nature}\ }\textbf {\bibinfo {volume} {425}},\ \bibinfo {pages}
  {941} (\bibinfo {year} {2003})}\BibitemShut {NoStop}%
\bibitem [{\citenamefont {McDermott}\ \emph {et~al.}(2005)\citenamefont
  {McDermott}, \citenamefont {Simmonds}, \citenamefont {Steffen}, \citenamefont
  {Cooper}, \citenamefont {Cicak}, \citenamefont {Osborn}, \citenamefont {Oh},
  \citenamefont {Pappas},\ and\ \citenamefont
  {Martinis}}]{SimulStateMeasureJosephsonQB}%
  \BibitemOpen
  \bibfield  {author} {\bibinfo {author} {\bibfnamefont {R.}~\bibnamefont
  {McDermott}}, \bibinfo {author} {\bibfnamefont {R.~W.}\ \bibnamefont
  {Simmonds}}, \bibinfo {author} {\bibfnamefont {M.}~\bibnamefont {Steffen}},
  \bibinfo {author} {\bibfnamefont {K.~B.}\ \bibnamefont {Cooper}}, \bibinfo
  {author} {\bibfnamefont {K.}~\bibnamefont {Cicak}}, \bibinfo {author}
  {\bibfnamefont {K.~D.}\ \bibnamefont {Osborn}}, \bibinfo {author}
  {\bibfnamefont {S.}~\bibnamefont {Oh}}, \bibinfo {author} {\bibfnamefont
  {D.~P.}\ \bibnamefont {Pappas}},\ and\ \bibinfo {author} {\bibfnamefont
  {J.~M.}\ \bibnamefont {Martinis}},\ }\href
  {https://doi.org/10.1126/science.1107572} {\bibfield  {journal} {\bibinfo
  {journal} {Science}\ }\textbf {\bibinfo {volume} {307}},\ \bibinfo {pages}
  {1299} (\bibinfo {year} {2005})}\BibitemShut {NoStop}%
\bibitem [{\citenamefont {Breuer}\ and\ \citenamefont
  {Petruccione}(2002)}]{TheoryOQSBook}%
  \BibitemOpen
  \bibfield  {author} {\bibinfo {author} {\bibfnamefont {H.~P.}\ \bibnamefont
  {Breuer}}\ and\ \bibinfo {author} {\bibfnamefont {F.}~\bibnamefont
  {Petruccione}},\ }\href@noop {} {\emph {\bibinfo {title} {The theory of open
  quantum systems}}}\ (\bibinfo  {publisher} {Oxford University Press},\
  \bibinfo {address} {Great Clarendon Street},\ \bibinfo {year}
  {2002})\BibitemShut {NoStop}%
\bibitem [{\citenamefont {Moy}\ \emph {et~al.}(1999)\citenamefont {Moy},
  \citenamefont {Hope},\ and\ \citenamefont
  {Savage}}]{BornMarkovApproxAtomLaser}%
  \BibitemOpen
  \bibfield  {author} {\bibinfo {author} {\bibfnamefont {G.~M.}\ \bibnamefont
  {Moy}}, \bibinfo {author} {\bibfnamefont {J.~J.}\ \bibnamefont {Hope}},\ and\
  \bibinfo {author} {\bibfnamefont {C.~M.}\ \bibnamefont {Savage}},\ }\href
  {https://doi.org/10.1103/PhysRevA.59.667} {\bibfield  {journal} {\bibinfo
  {journal} {Phys. Rev. A}\ }\textbf {\bibinfo {volume} {59}},\ \bibinfo
  {pages} {667} (\bibinfo {year} {1999})}\BibitemShut {NoStop}%
\bibitem [{\citenamefont {Burgess}\ and\ \citenamefont
  {Florescu}(2021)}]{burgess_modelling_2021}%
  \BibitemOpen
  \bibfield  {author} {\bibinfo {author} {\bibfnamefont {A.}~\bibnamefont
  {Burgess}}\ and\ \bibinfo {author} {\bibfnamefont {M.}~\bibnamefont
  {Florescu}},\ }\href {https://doi.org/10.1364/OME.425263} {\bibfield
  {journal} {\bibinfo  {journal} {Opt. Mat. Express}\ }\textbf {\bibinfo
  {volume} {11}},\ \bibinfo {pages} {2037} (\bibinfo {year}
  {2021})}\BibitemShut {NoStop}%
\bibitem [{\citenamefont {Florescu}\ and\ \citenamefont
  {John}(2001)}]{SingleAtomSwitch}%
  \BibitemOpen
  \bibfield  {author} {\bibinfo {author} {\bibfnamefont {M.}~\bibnamefont
  {Florescu}}\ and\ \bibinfo {author} {\bibfnamefont {S.}~\bibnamefont
  {John}},\ }\href {https://doi.org/10.1103/PhysRevA.64.033801} {\bibfield
  {journal} {\bibinfo  {journal} {Phys. Rev. A}\ }\textbf {\bibinfo {volume}
  {64}},\ \bibinfo {pages} {033801} (\bibinfo {year} {2001})}\BibitemShut
  {NoStop}%
\bibitem [{\citenamefont {John}\ and\ \citenamefont
  {Florescu}(2001)}]{john_florescu_2001}%
  \BibitemOpen
  \bibfield  {author} {\bibinfo {author} {\bibfnamefont {S.}~\bibnamefont
  {John}}\ and\ \bibinfo {author} {\bibfnamefont {M.}~\bibnamefont
  {Florescu}},\ }\bibfield  {journal} {\bibinfo  {journal} {Journal of Optics
  A: Pure and Applied Optics}\ }\textbf {\bibinfo {volume} {3}},\ \href
  {https://doi.org/10.1088/1464-4258/3/6/361} {10.1088/1464-4258/3/6/361}
  (\bibinfo {year} {2001})\BibitemShut {NoStop}%
\bibitem [{\citenamefont {John}\ and\ \citenamefont {Quang}(1994)}]{John1994}%
  \BibitemOpen
  \bibfield  {author} {\bibinfo {author} {\bibfnamefont {S.}~\bibnamefont
  {John}}\ and\ \bibinfo {author} {\bibfnamefont {T.}~\bibnamefont {Quang}},\
  }\href {https://doi.org/10.1103/physreva.50.1764} {\bibfield  {journal}
  {\bibinfo  {journal} {Physical Review A}\ }\textbf {\bibinfo {volume} {50}},\
  \bibinfo {pages} {1764} (\bibinfo {year} {1994})}\BibitemShut {NoStop}%
\bibitem [{\citenamefont {Kimble}(2008)}]{KimbleQuantumNetwork}%
  \BibitemOpen
  \bibfield  {author} {\bibinfo {author} {\bibfnamefont {H.~J.}\ \bibnamefont
  {Kimble}},\ }\href {https://doi.org/10.1038/nature07127} {\bibfield
  {journal} {\bibinfo  {journal} {Nature}\ }\textbf {\bibinfo {volume} {453}},\
  \bibinfo {pages} {1023} (\bibinfo {year} {2008})}\BibitemShut {NoStop}%
\bibitem [{\citenamefont {K{\'o}m{\'a}r}\ \emph {et~al.}(2014)\citenamefont
  {K{\'o}m{\'a}r}, \citenamefont {Kessler}, \citenamefont {Bishof},
  \citenamefont {Jiang}, \citenamefont {S{\o}rensen}, \citenamefont {Ye},\ and\
  \citenamefont {Lukin}}]{QuantumNetworkClocks}%
  \BibitemOpen
  \bibfield  {author} {\bibinfo {author} {\bibfnamefont {P.}~\bibnamefont
  {K{\'o}m{\'a}r}}, \bibinfo {author} {\bibfnamefont {E.~M.}\ \bibnamefont
  {Kessler}}, \bibinfo {author} {\bibfnamefont {M.}~\bibnamefont {Bishof}},
  \bibinfo {author} {\bibfnamefont {L.}~\bibnamefont {Jiang}}, \bibinfo
  {author} {\bibfnamefont {A.~S.}\ \bibnamefont {S{\o}rensen}}, \bibinfo
  {author} {\bibfnamefont {J.}~\bibnamefont {Ye}},\ and\ \bibinfo {author}
  {\bibfnamefont {M.~D.}\ \bibnamefont {Lukin}},\ }\href
  {https://doi.org/10.1038/nphys3000} {\bibfield  {journal} {\bibinfo
  {journal} {Nature Physics}\ }\textbf {\bibinfo {volume} {10}},\ \bibinfo
  {pages} {582} (\bibinfo {year} {2014})}\BibitemShut {NoStop}%
\bibitem [{\citenamefont {Schliemann}\ \emph {et~al.}(2003)\citenamefont
  {Schliemann}, \citenamefont {Khaetskii},\ and\ \citenamefont
  {Loss}}]{quantumdotspincouple}%
  \BibitemOpen
  \bibfield  {author} {\bibinfo {author} {\bibfnamefont {J.}~\bibnamefont
  {Schliemann}}, \bibinfo {author} {\bibfnamefont {A.}~\bibnamefont
  {Khaetskii}},\ and\ \bibinfo {author} {\bibfnamefont {D.}~\bibnamefont
  {Loss}},\ }\href {https://doi.org/10.1088/0953-8984/15/50/r01} {\bibfield
  {journal} {\bibinfo  {journal} {Journal of Physics: Condensed Matter}\
  }\textbf {\bibinfo {volume} {15}},\ \bibinfo {pages} {R1809} (\bibinfo {year}
  {2003})}\BibitemShut {NoStop}%
\bibitem [{\citenamefont {Fazio}\ and\ \citenamefont {{van der
  Zant}}(2001)}]{Superconducting}%
  \BibitemOpen
  \bibfield  {author} {\bibinfo {author} {\bibfnamefont {R.}~\bibnamefont
  {Fazio}}\ and\ \bibinfo {author} {\bibfnamefont {H.}~\bibnamefont {{van der
  Zant}}},\ }\href
  {https://doi.org/https://doi.org/10.1016/S0370-1573(01)00022-9} {\bibfield
  {journal} {\bibinfo  {journal} {Physics Reports}\ }\textbf {\bibinfo {volume}
  {355}},\ \bibinfo {pages} {235} (\bibinfo {year} {2001})}\BibitemShut
  {NoStop}%
\bibitem [{\citenamefont {Porras}\ and\ \citenamefont
  {Cirac}(2004)}]{trappedion}%
  \BibitemOpen
  \bibfield  {author} {\bibinfo {author} {\bibfnamefont {D.}~\bibnamefont
  {Porras}}\ and\ \bibinfo {author} {\bibfnamefont {J.~I.}\ \bibnamefont
  {Cirac}},\ }\href {https://doi.org/10.1103/PhysRevLett.92.207901} {\bibfield
  {journal} {\bibinfo  {journal} {Phys. Rev. Lett.}\ }\textbf {\bibinfo
  {volume} {92}},\ \bibinfo {pages} {207901} (\bibinfo {year}
  {2004})}\BibitemShut {NoStop}%
\bibitem [{\citenamefont {Duan}\ \emph {et~al.}(2003)\citenamefont {Duan},
  \citenamefont {Demler},\ and\ \citenamefont {Lukin}}]{OpticalLattice}%
  \BibitemOpen
  \bibfield  {author} {\bibinfo {author} {\bibfnamefont {L.-M.}\ \bibnamefont
  {Duan}}, \bibinfo {author} {\bibfnamefont {E.}~\bibnamefont {Demler}},\ and\
  \bibinfo {author} {\bibfnamefont {M.~D.}\ \bibnamefont {Lukin}},\ }\href
  {https://doi.org/10.1103/PhysRevLett.91.090402} {\bibfield  {journal}
  {\bibinfo  {journal} {Phys. Rev. Lett.}\ }\textbf {\bibinfo {volume} {91}},\
  \bibinfo {pages} {090402} (\bibinfo {year} {2003})}\BibitemShut {NoStop}%
\bibitem [{\citenamefont {Hood}\ \emph
  {et~al.}(2016{\natexlab{a}})\citenamefont {Hood}, \citenamefont {Goban},
  \citenamefont {Asenjo-Garcia}, \citenamefont {Lu}, \citenamefont {Yu},
  \citenamefont {Chang},\ and\ \citenamefont {Kimble}}]{KimbleOPhCr}%
  \BibitemOpen
  \bibfield  {author} {\bibinfo {author} {\bibfnamefont {J.~D.}\ \bibnamefont
  {Hood}}, \bibinfo {author} {\bibfnamefont {A.}~\bibnamefont {Goban}},
  \bibinfo {author} {\bibfnamefont {A.}~\bibnamefont {Asenjo-Garcia}}, \bibinfo
  {author} {\bibfnamefont {M.}~\bibnamefont {Lu}}, \bibinfo {author}
  {\bibfnamefont {S.-P.}\ \bibnamefont {Yu}}, \bibinfo {author} {\bibfnamefont
  {D.~E.}\ \bibnamefont {Chang}},\ and\ \bibinfo {author} {\bibfnamefont
  {H.~J.}\ \bibnamefont {Kimble}},\ }\href
  {https://doi.org/10.1073/pnas.1603788113} {\bibfield  {journal} {\bibinfo
  {journal} {Proceedings of the National Academy of Sciences}\ }\textbf
  {\bibinfo {volume} {113}},\ \bibinfo {pages} {10507} (\bibinfo {year}
  {2016}{\natexlab{a}})},\ \Eprint
  {https://arxiv.org/abs/https://www.pnas.org/content/113/38/10507.full.pdf}
  {https://www.pnas.org/content/113/38/10507.full.pdf} \BibitemShut {NoStop}%
\bibitem [{\citenamefont {Yu}\ \emph {et~al.}(2019)\citenamefont {Yu},
  \citenamefont {Muniz}, \citenamefont {Hung},\ and\ \citenamefont
  {Kimble}}]{Yu12743}%
  \BibitemOpen
  \bibfield  {author} {\bibinfo {author} {\bibfnamefont {S.-P.}\ \bibnamefont
  {Yu}}, \bibinfo {author} {\bibfnamefont {J.~A.}\ \bibnamefont {Muniz}},
  \bibinfo {author} {\bibfnamefont {C.-L.}\ \bibnamefont {Hung}},\ and\
  \bibinfo {author} {\bibfnamefont {H.~J.}\ \bibnamefont {Kimble}},\ }\href
  {https://doi.org/10.1073/pnas.1822110116} {\bibfield  {journal} {\bibinfo
  {journal} {Proceedings of the National Academy of Sciences}\ }\textbf
  {\bibinfo {volume} {116}},\ \bibinfo {pages} {12743} (\bibinfo {year}
  {2019})},\ \Eprint
  {https://arxiv.org/abs/https://www.pnas.org/content/116/26/12743.full.pdf}
  {https://www.pnas.org/content/116/26/12743.full.pdf} \BibitemShut {NoStop}%
\bibitem [{\citenamefont {Javadi}\ \emph {et~al.}(2015)\citenamefont {Javadi},
  \citenamefont {S{\"o}llner}, \citenamefont {Arcari}, \citenamefont {Hansen},
  \citenamefont {Midolo}, \citenamefont {Mahmoodian}, \citenamefont
  {Kir{\v{s}}ansk{\.{e}}}, \citenamefont {Pregnolato}, \citenamefont {Lee},
  \citenamefont {Song}, \citenamefont {Stobbe},\ and\ \citenamefont
  {Lodahl}}]{SinglePhotonQD}%
  \BibitemOpen
  \bibfield  {author} {\bibinfo {author} {\bibfnamefont {A.}~\bibnamefont
  {Javadi}}, \bibinfo {author} {\bibfnamefont {I.}~\bibnamefont {S{\"o}llner}},
  \bibinfo {author} {\bibfnamefont {M.}~\bibnamefont {Arcari}}, \bibinfo
  {author} {\bibfnamefont {S.~L.}\ \bibnamefont {Hansen}}, \bibinfo {author}
  {\bibfnamefont {L.}~\bibnamefont {Midolo}}, \bibinfo {author} {\bibfnamefont
  {S.}~\bibnamefont {Mahmoodian}}, \bibinfo {author} {\bibfnamefont
  {G.}~\bibnamefont {Kir{\v{s}}ansk{\.{e}}}}, \bibinfo {author} {\bibfnamefont
  {T.}~\bibnamefont {Pregnolato}}, \bibinfo {author} {\bibfnamefont {E.~H.}\
  \bibnamefont {Lee}}, \bibinfo {author} {\bibfnamefont {J.~D.}\ \bibnamefont
  {Song}}, \bibinfo {author} {\bibfnamefont {S.}~\bibnamefont {Stobbe}},\ and\
  \bibinfo {author} {\bibfnamefont {P.}~\bibnamefont {Lodahl}},\ }\href
  {https://doi.org/10.1038/ncomms9655} {\bibfield  {journal} {\bibinfo
  {journal} {Nature Communications}\ }\textbf {\bibinfo {volume} {6}},\
  \bibinfo {pages} {8655} (\bibinfo {year} {2015})}\BibitemShut {NoStop}%
\bibitem [{\citenamefont {Sinayskiy}\ \emph {et~al.}(2009)\citenamefont
  {Sinayskiy}, \citenamefont {Ferraro}, \citenamefont {Napoli}, \citenamefont
  {Messina},\ and\ \citenamefont {Petruccione}}]{NM2Qubit}%
  \BibitemOpen
  \bibfield  {author} {\bibinfo {author} {\bibfnamefont {I.}~\bibnamefont
  {Sinayskiy}}, \bibinfo {author} {\bibfnamefont {E.}~\bibnamefont {Ferraro}},
  \bibinfo {author} {\bibfnamefont {A.}~\bibnamefont {Napoli}}, \bibinfo
  {author} {\bibfnamefont {A.}~\bibnamefont {Messina}},\ and\ \bibinfo {author}
  {\bibfnamefont {F.}~\bibnamefont {Petruccione}},\ }\href
  {https://doi.org/10.1088/1751-8113/42/48/485301} {\bibfield  {journal}
  {\bibinfo  {journal} {Journal of Physics A: Mathematical and Theoretical}\
  }\textbf {\bibinfo {volume} {42}},\ \bibinfo {pages} {485301} (\bibinfo
  {year} {2009})}\BibitemShut {NoStop}%
\bibitem [{\citenamefont {Cui}\ \emph {et~al.}(2009)\citenamefont {Cui},
  \citenamefont {Xi},\ and\ \citenamefont {Pan}}]{NM2qubitSE}%
  \BibitemOpen
  \bibfield  {author} {\bibinfo {author} {\bibfnamefont {W.}~\bibnamefont
  {Cui}}, \bibinfo {author} {\bibfnamefont {Z.}~\bibnamefont {Xi}},\ and\
  \bibinfo {author} {\bibfnamefont {Y.}~\bibnamefont {Pan}},\ }\href
  {https://doi.org/10.1088/1751-8113/42/15/155303} {\bibfield  {journal}
  {\bibinfo  {journal} {Journal of Physics A: Mathematical and Theoretical}\
  }\textbf {\bibinfo {volume} {42}},\ \bibinfo {pages} {155303} (\bibinfo
  {year} {2009})}\BibitemShut {NoStop}%
\bibitem [{\citenamefont {Breuer}\ \emph {et~al.}(1999)\citenamefont {Breuer},
  \citenamefont {Kappler},\ and\ \citenamefont {Petruccione}}]{Breuer_1999}%
  \BibitemOpen
  \bibfield  {author} {\bibinfo {author} {\bibfnamefont {H.-P.}\ \bibnamefont
  {Breuer}}, \bibinfo {author} {\bibfnamefont {B.}~\bibnamefont {Kappler}},\
  and\ \bibinfo {author} {\bibfnamefont {F.}~\bibnamefont {Petruccione}},\
  }\href {https://doi.org/10.1103/physreva.59.1633} {\bibfield  {journal}
  {\bibinfo  {journal} {Physical Review A}\ }\textbf {\bibinfo {volume} {59}},\
  \bibinfo {pages} {1633–1643} (\bibinfo {year} {1999})}\BibitemShut
  {NoStop}%
\bibitem [{\citenamefont {Shen}\ \emph {et~al.}(2016)\citenamefont {Shen},
  \citenamefont {Shao}, \citenamefont {Wang}, \citenamefont {Zhao},\ and\
  \citenamefont {Yi}}]{shen_quantum_2016}%
  \BibitemOpen
  \bibfield  {author} {\bibinfo {author} {\bibfnamefont {H.~Z.}\ \bibnamefont
  {Shen}}, \bibinfo {author} {\bibfnamefont {X.~Q.}\ \bibnamefont {Shao}},
  \bibinfo {author} {\bibfnamefont {G.~C.}\ \bibnamefont {Wang}}, \bibinfo
  {author} {\bibfnamefont {X.~L.}\ \bibnamefont {Zhao}},\ and\ \bibinfo
  {author} {\bibfnamefont {X.~X.}\ \bibnamefont {Yi}},\ }\href
  {https://doi.org/10.1103/PhysRevE.93.012107} {\bibfield  {journal} {\bibinfo
  {journal} {Physical Review E}\ }\textbf {\bibinfo {volume} {93}},\ \bibinfo
  {pages} {012107} (\bibinfo {year} {2016})}\BibitemShut {NoStop}%
\bibitem [{\citenamefont {Ho}\ \emph {et~al.}(1990)\citenamefont {Ho},
  \citenamefont {Chan},\ and\ \citenamefont {Soukoulis}}]{Existence}%
  \BibitemOpen
  \bibfield  {author} {\bibinfo {author} {\bibfnamefont {K.~M.}\ \bibnamefont
  {Ho}}, \bibinfo {author} {\bibfnamefont {C.~T.}\ \bibnamefont {Chan}},\ and\
  \bibinfo {author} {\bibfnamefont {C.~M.}\ \bibnamefont {Soukoulis}},\ }\href
  {https://doi.org/10.1103/PhysRevLett.65.3152} {\bibfield  {journal} {\bibinfo
   {journal} {Phys. Rev. Lett.}\ }\textbf {\bibinfo {volume} {65}},\ \bibinfo
  {pages} {3152} (\bibinfo {year} {1990})}\BibitemShut {NoStop}%
\bibitem [{\citenamefont {Yablonovitch}\ \emph {et~al.}(1991)\citenamefont
  {Yablonovitch}, \citenamefont {Gmitter},\ and\ \citenamefont {Leung}}]{FCC}%
  \BibitemOpen
  \bibfield  {author} {\bibinfo {author} {\bibfnamefont {E.}~\bibnamefont
  {Yablonovitch}}, \bibinfo {author} {\bibfnamefont {T.~J.}\ \bibnamefont
  {Gmitter}},\ and\ \bibinfo {author} {\bibfnamefont {K.~M.}\ \bibnamefont
  {Leung}},\ }\href {https://doi.org/10.1103/PhysRevLett.67.2295} {\bibfield
  {journal} {\bibinfo  {journal} {Phys. Rev. Lett.}\ }\textbf {\bibinfo
  {volume} {67}},\ \bibinfo {pages} {2295} (\bibinfo {year}
  {1991})}\BibitemShut {NoStop}%
\bibitem [{\citenamefont {John}\ and\ \citenamefont
  {Quang}(1995)}]{SuperRadiancePBG}%
  \BibitemOpen
  \bibfield  {author} {\bibinfo {author} {\bibfnamefont {S.}~\bibnamefont
  {John}}\ and\ \bibinfo {author} {\bibfnamefont {T.}~\bibnamefont {Quang}},\
  }\href {https://doi.org/10.1103/PhysRevLett.74.3419} {\bibfield  {journal}
  {\bibinfo  {journal} {Phys. Rev. Lett.}\ }\textbf {\bibinfo {volume} {74}},\
  \bibinfo {pages} {3419} (\bibinfo {year} {1995})}\BibitemShut {NoStop}%
\bibitem [{\citenamefont {Gross}\ and\ \citenamefont
  {Haroche}(1982)}]{Superradiance}%
  \BibitemOpen
  \bibfield  {author} {\bibinfo {author} {\bibfnamefont {M.}~\bibnamefont
  {Gross}}\ and\ \bibinfo {author} {\bibfnamefont {S.}~\bibnamefont
  {Haroche}},\ }\href
  {https://doi.org/https://doi.org/10.1016/0370-1573(82)90102-8} {\bibfield
  {journal} {\bibinfo  {journal} {Physics Reports}\ }\textbf {\bibinfo {volume}
  {93}},\ \bibinfo {pages} {301} (\bibinfo {year} {1982})}\BibitemShut
  {NoStop}%
\bibitem [{\citenamefont {Dicke}(1954)}]{Dicke}%
  \BibitemOpen
  \bibfield  {author} {\bibinfo {author} {\bibfnamefont {R.~H.}\ \bibnamefont
  {Dicke}},\ }\href {https://doi.org/10.1103/PhysRev.93.99} {\bibfield
  {journal} {\bibinfo  {journal} {Phys. Rev.}\ }\textbf {\bibinfo {volume}
  {93}},\ \bibinfo {pages} {99} (\bibinfo {year} {1954})}\BibitemShut {NoStop}%
\bibitem [{\citenamefont {Pfeifer}(1982)}]{Pfeifer}%
  \BibitemOpen
  \bibfield  {author} {\bibinfo {author} {\bibfnamefont {P.}~\bibnamefont
  {Pfeifer}},\ }\href {https://doi.org/10.1103/PhysRevA.26.701} {\bibfield
  {journal} {\bibinfo  {journal} {Phys. Rev. A}\ }\textbf {\bibinfo {volume}
  {26}},\ \bibinfo {pages} {701} (\bibinfo {year} {1982})}\BibitemShut
  {NoStop}%
\bibitem [{\citenamefont {Zienau}(1975)}]{AllenEberly}%
  \BibitemOpen
  \bibfield  {author} {\bibinfo {author} {\bibfnamefont {S.}~\bibnamefont
  {Zienau}},\ }\href {https://doi.org/10.1088/0031-9112/26/12/039} {\bibfield
  {journal} {\bibinfo  {journal} {Physics Bulletin}\ }\textbf {\bibinfo
  {volume} {26}},\ \bibinfo {pages} {545} (\bibinfo {year} {1975})}\BibitemShut
  {NoStop}%
\bibitem [{\citenamefont {Garraway}(1997)}]{Garraway}%
  \BibitemOpen
  \bibfield  {author} {\bibinfo {author} {\bibfnamefont {B.~M.}\ \bibnamefont
  {Garraway}},\ }\href {https://doi.org/10.1103/PhysRevA.55.2290} {\bibfield
  {journal} {\bibinfo  {journal} {Phys. Rev. A}\ }\textbf {\bibinfo {volume}
  {55}},\ \bibinfo {pages} {2290} (\bibinfo {year} {1997})}\BibitemShut
  {NoStop}%
\bibitem [{\citenamefont {Plemelj}(1908)}]{plemelj1908erganzungssatz}%
  \BibitemOpen
  \bibfield  {author} {\bibinfo {author} {\bibfnamefont {J.}~\bibnamefont
  {Plemelj}},\ }\href@noop {} {\bibfield  {journal} {\bibinfo  {journal}
  {Monatshefte f{\"u}r Mathematik und Physik}\ }\textbf {\bibinfo {volume}
  {19}},\ \bibinfo {pages} {205} (\bibinfo {year} {1908})}\BibitemShut
  {NoStop}%
\bibitem [{\citenamefont {Davis}(1959)}]{Gamma}%
  \BibitemOpen
  \bibfield  {author} {\bibinfo {author} {\bibfnamefont {P.~J.}\ \bibnamefont
  {Davis}},\ }\href {http://www.jstor.org/stable/2309786} {\bibfield  {journal}
  {\bibinfo  {journal} {The American Mathematical Monthly}\ }\textbf {\bibinfo
  {volume} {66}},\ \bibinfo {pages} {849} (\bibinfo {year} {1959})}\BibitemShut
  {NoStop}%
\bibitem [{{\relax DLMF}()}]{IncompleteGamma}%
  \BibitemOpen
  {\relax DLMF},\ \href {https://dlmf.nist.gov/8.2} {\bibinfo {title} {{\it
  NIST Digital Library of Mathematical Functions}}},\ \bibinfo {howpublished}
  {https://dlmf.nist.gov/8.2, Release 1.1.2 of 2021-06-15} (\bibinfo {year}
  {2021}),\ \bibinfo {note} {f.~W.~J. Olver, A.~B. {Olde Daalhuis}, D.~W.
  Lozier, B.~I. Schneider, R.~F. Boisvert, C.~W. Clark, B.~R. Miller, B.~V.
  Saunders, H.~S. Cohl, and M.~A. McClain, eds.}\BibitemShut {Stop}%
\bibitem [{\citenamefont {Goban}\ \emph {et~al.}(2015)\citenamefont {Goban},
  \citenamefont {Hung}, \citenamefont {Hood}, \citenamefont {Yu}, \citenamefont
  {Muniz}, \citenamefont {Painter},\ and\ \citenamefont
  {Kimble}}]{KimbleSuperradiance}%
  \BibitemOpen
  \bibfield  {author} {\bibinfo {author} {\bibfnamefont {A.}~\bibnamefont
  {Goban}}, \bibinfo {author} {\bibfnamefont {C.-L.}\ \bibnamefont {Hung}},
  \bibinfo {author} {\bibfnamefont {J.~D.}\ \bibnamefont {Hood}}, \bibinfo
  {author} {\bibfnamefont {S.-P.}\ \bibnamefont {Yu}}, \bibinfo {author}
  {\bibfnamefont {J.~A.}\ \bibnamefont {Muniz}}, \bibinfo {author}
  {\bibfnamefont {O.}~\bibnamefont {Painter}},\ and\ \bibinfo {author}
  {\bibfnamefont {H.~J.}\ \bibnamefont {Kimble}},\ }\href
  {https://doi.org/10.1103/PhysRevLett.115.063601} {\bibfield  {journal}
  {\bibinfo  {journal} {Phys. Rev. Lett.}\ }\textbf {\bibinfo {volume} {115}},\
  \bibinfo {pages} {063601} (\bibinfo {year} {2015})}\BibitemShut {NoStop}%
\bibitem [{\citenamefont {Kim}\ \emph {et~al.}(2016)\citenamefont {Kim},
  \citenamefont {Kim}, \citenamefont {Kyhm}, \citenamefont {Taylor},
  \citenamefont {Kim}, \citenamefont {Song}, \citenamefont {Je},\ and\
  \citenamefont {Dang}}]{dipole-dipole}%
  \BibitemOpen
  \bibfield  {author} {\bibinfo {author} {\bibfnamefont {H.}~\bibnamefont
  {Kim}}, \bibinfo {author} {\bibfnamefont {I.}~\bibnamefont {Kim}}, \bibinfo
  {author} {\bibfnamefont {K.}~\bibnamefont {Kyhm}}, \bibinfo {author}
  {\bibfnamefont {R.~A.}\ \bibnamefont {Taylor}}, \bibinfo {author}
  {\bibfnamefont {J.~S.}\ \bibnamefont {Kim}}, \bibinfo {author} {\bibfnamefont
  {J.~D.}\ \bibnamefont {Song}}, \bibinfo {author} {\bibfnamefont {K.~C.}\
  \bibnamefont {Je}},\ and\ \bibinfo {author} {\bibfnamefont {L.~S.}\
  \bibnamefont {Dang}},\ }\href {https://doi.org/10.1021/acs.nanolett.6b03868}
  {\bibfield  {journal} {\bibinfo  {journal} {Nano Letters}\ }\textbf {\bibinfo
  {volume} {16}},\ \bibinfo {pages} {7755} (\bibinfo {year} {2016})},\ \bibinfo
  {note} {pMID: 27960477},\ \Eprint
  {https://arxiv.org/abs/https://doi.org/10.1021/acs.nanolett.6b03868}
  {https://doi.org/10.1021/acs.nanolett.6b03868} \BibitemShut {NoStop}%
\bibitem [{\citenamefont {Pach{\'{o}}n}\ \emph {et~al.}(2017)\citenamefont
  {Pach{\'{o}}n}, \citenamefont {Botero},\ and\ \citenamefont {Brumer}}]{LHC}%
  \BibitemOpen
  \bibfield  {author} {\bibinfo {author} {\bibfnamefont {L.~A.}\ \bibnamefont
  {Pach{\'{o}}n}}, \bibinfo {author} {\bibfnamefont {J.~D.}\ \bibnamefont
  {Botero}},\ and\ \bibinfo {author} {\bibfnamefont {P.}~\bibnamefont
  {Brumer}},\ }\href {https://doi.org/10.1088/1361-6455/aa8696} {\bibfield
  {journal} {\bibinfo  {journal} {Journal of Physics B: Atomic, Molecular and
  Optical Physics}\ }\textbf {\bibinfo {volume} {50}},\ \bibinfo {pages}
  {184003} (\bibinfo {year} {2017})}\BibitemShut {NoStop}%
\bibitem [{\citenamefont {John}\ and\ \citenamefont {Wang}(1990)}]{QEDNearPBG}%
  \BibitemOpen
  \bibfield  {author} {\bibinfo {author} {\bibfnamefont {S.}~\bibnamefont
  {John}}\ and\ \bibinfo {author} {\bibfnamefont {J.}~\bibnamefont {Wang}},\
  }\href {https://doi.org/10.1103/PhysRevLett.64.2418} {\bibfield  {journal}
  {\bibinfo  {journal} {Phys. Rev. Lett.}\ }\textbf {\bibinfo {volume} {64}},\
  \bibinfo {pages} {2418} (\bibinfo {year} {1990})}\BibitemShut {NoStop}%
\bibitem [{\citenamefont {Dutta}\ \emph {et~al.}(2016)\citenamefont {Dutta},
  \citenamefont {Goyal}, \citenamefont {Srivastava},\ and\ \citenamefont
  {Pal}}]{PCWaveguide}%
  \BibitemOpen
  \bibfield  {author} {\bibinfo {author} {\bibfnamefont {H.~S.}\ \bibnamefont
  {Dutta}}, \bibinfo {author} {\bibfnamefont {A.~K.}\ \bibnamefont {Goyal}},
  \bibinfo {author} {\bibfnamefont {V.}~\bibnamefont {Srivastava}},\ and\
  \bibinfo {author} {\bibfnamefont {S.}~\bibnamefont {Pal}},\ }\href
  {https://doi.org/https://doi.org/10.1016/j.photonics.2016.04.001} {\bibfield
  {journal} {\bibinfo  {journal} {Photonics and Nanostructures - Fundamentals
  and Applications}\ }\textbf {\bibinfo {volume} {20}},\ \bibinfo {pages} {41}
  (\bibinfo {year} {2016})}\BibitemShut {NoStop}%
\bibitem [{\citenamefont {Hood}\ \emph
  {et~al.}(2016{\natexlab{b}})\citenamefont {Hood}, \citenamefont {Goban},
  \citenamefont {Asenjo-Garcia}, \citenamefont {Lu}, \citenamefont {Yu},
  \citenamefont {Chang},\ and\ \citenamefont {Kimble}}]{ManyAtomsInPCW}%
  \BibitemOpen
  \bibfield  {author} {\bibinfo {author} {\bibfnamefont {J.~D.}\ \bibnamefont
  {Hood}}, \bibinfo {author} {\bibfnamefont {A.}~\bibnamefont {Goban}},
  \bibinfo {author} {\bibfnamefont {A.}~\bibnamefont {Asenjo-Garcia}}, \bibinfo
  {author} {\bibfnamefont {M.}~\bibnamefont {Lu}}, \bibinfo {author}
  {\bibfnamefont {S.-P.}\ \bibnamefont {Yu}}, \bibinfo {author} {\bibfnamefont
  {D.~E.}\ \bibnamefont {Chang}},\ and\ \bibinfo {author} {\bibfnamefont
  {H.~J.}\ \bibnamefont {Kimble}},\ }\href
  {https://doi.org/10.1073/pnas.1603788113} {\bibfield  {journal} {\bibinfo
  {journal} {Proceedings of the National Academy of Sciences}\ }\textbf
  {\bibinfo {volume} {113}},\ \bibinfo {pages} {10507} (\bibinfo {year}
  {2016}{\natexlab{b}})},\ \Eprint
  {https://arxiv.org/abs/https://www.pnas.org/content/113/38/10507.full.pdf}
  {https://www.pnas.org/content/113/38/10507.full.pdf} \BibitemShut {NoStop}%
\end{thebibliography}%
\end{document}